\documentclass{article}

\usepackage{epsfig}
\usepackage{graphics}
\usepackage{graphicx}
\usepackage[centertags]{amsmath}
\usepackage{amsfonts}
\usepackage{amsthm}
\usepackage{amssymb}

\usepackage{url}
\usepackage{subfigure}

\usepackage{color}
\usepackage{verbatim}

\begin{document}

\title{How to make dull cellular automata complex by adding memory: Rule 126 case study}

\author{Genaro J. Mart\'{\i}nez$^{1,2}$, Andrew Adamatzky$^2$ \\ Juan C. Seck-Tuoh-Mora$^3$, Ramon Alonso-Sanz$^{2,4}$}

\date{June, 2009}

\maketitle

\begin{centering}
$^1$ Instituto de Ciencias Nucleares and Centro de Ciencias de la Complejidad, Universidad Nacional Aut\'onoma de M\'exico, M\'exico. \\
\url{genaro.martinez@uwe.ac.uk} \\

$^2$ Bristol Institute of Technology, University of the West of England, Bristol, United Kingdom. \\
\url{andrew.adamatzky@uwe.ac.uk} \\

$^3$ Centro de Investigaci\'on Avanzada en Ingenier\'{\i}a Industrial, Universidad Aut\'onoma del Estado de Hidalgo Pachuca, Hidalgo, M\'exico. \\
\url{jseck@uaeh.edu.mx} \\

$^4$ ETSI Agr\'onomos, Polytechnic University of Madrid, Madrid, Spain. \\
\url{ramon.alonso@upm.es} \\

\vspace{0.5cm}

{\small  \bf
Published in \\
Mart'nez~G.~J., Adamatzky~A., Seck-Tuoh-Mora~J..C., Alonso-Sanz~R.\\
How to make dull cellular automata complex by adding memory:\\ Rule 126 case study. 
Complexity 15 (2010) 6, 34--49.\\

\url{http://onlinelibrary.wiley.com/doi/10.1002/cplx.20311/abstract}
}
\end{centering}

\begin{abstract}
Using Rule 126 elementary cellular automaton (ECA) we demonstrate that a chaotic discrete system --- when enriched with memory -- hence exhibits complex dynamics where such space exploits on an ample universe of periodic patterns induced from original information of the ahistorical system. First we analyse classic ECA Rule 126 to identify basic characteristics with mean field theory, basins, and de Bruijn diagrams. In order to derive this complex dynamics, we use a kind of {\it memory} on Rule 126; from here interactions between gliders are studied for detecting stationary patterns, glider guns and simulating specific simple computable functions produced by glider collisions.
\vspace{0.5cm}

\end{abstract}

\newpage

%%%%%%%%%%%%%%%%%%%%%%
\section{Introduction}
In this paper we are making use of the memory tool to get a complex system from a chaotic function in discrete dynamical environments. Such technique takes the past history of the system for constructing its present and future: the {\it memory} \cite{kn:AMM01, kn:AM02, kn:Alo03, kn:AM03, kn:Alo06, kn:Alo09}. It was previously reported in \cite{kn:MAA09} how the chaotic ECA Rule 30 is decomposed into a complex system applying memory on this function. Recent results show that other chaotic functions (Rule 86 and Rule 101) yield complex dynamics selecting a kind of memory, including a controller to obtain self-organization by structure reactions and simple computations implemented by soliton reactions \cite{kn:MAA09a}.

We focus this work on cellular automata (CA) evolving in one dimension, in particular taking the well-known ECA where each function evaluates a central cell and its two nearest neighbors (from left and right) and, every cell takes a value from a binary alphabet. Such ECA were introduced by Wolfram and have been widely studied in several directions \cite{kn:Wolf94}.

Among such ECA, there is a set of functions evolving in chaotic global behaviour where a number of cells remain unordered and their transitions have a large number of ancestors \cite{kn:Wue94}. In this sense, special attention is given on a chaotic one: the ECA Rule 126 introduced by Wolfram in \cite{kn:Wolf83}. Particularly in \cite{kn:Wolf02},\footnote{http://mathworld.wolfram.com/Rule126.html} it is commented that Rule 126 generates a regular language with average growing faster than any polynomial time. This property can be analysed as well by de Bruijn diagrams with additional features; in particular another partial analysis with de Bruijn and subset diagrams was done by McIntosh \cite{kn:Mc09} showing that string 010 in Rule 126 is the minimal {\it Garden of Eden} configuration (having no ancestors).

Based on the idea developed in previous results for obtaining complex dynamics from chaotic functions selecting memory and working systematically, it was suspected that a complex dynamics may emerge in Rule 126 given its relation to regular languages; making use of  gliders coded by regular expressions, as it was studied in Rule 110 \cite{kn:MMS08} and Rule 54 \cite{kn:MAM08}.

In this way Rule 126 provides a special case of how a chaotic behaviour can be decomposed selecting a kind of memory into a extraordinary activity of gliders, glider guns, still-life structures and a huge number of reactions. Such features can be compared to Brain Brian's rule behaviour or Conway's Life but in one dimension; actually none traditional ECA could have a glider dynamics comparable to the one revealed in this ECA with memory denoted as $\phi_{R126maj}$ (following notation described in \cite{kn:MAA09, kn:MAA09a}).

%%%%%%%%%%%%%%%%%%%%%%
\subsection{One-dimensional cellular automata}
One-dimensional CA is represented by an array of {\it cells} $x_i$ where $i \in \mathbb{Z}$ (integer set) and each $x$ takes a value from a finite alphabet $\Sigma$. Thus, a sequence of cells \{$x_i$\} of finite length $n$ describes a string or {\it global configuration} $c$ on $\Sigma$. This way, the set of finite configurations will be expressed as $\Sigma^n$. An evolution is comprised by a sequence of configurations $\{c_i\}$ produced by the mapping $\Phi:\Sigma^n \rightarrow \Sigma^n$; thus the global relation is symbolized as:

\begin{equation}
\Phi(c^t) \rightarrow c^{t+1}
\label{globalFunction}
\end{equation}

\noindent where $t$ represents time and every global state of $c$ is defined by a sequence of cell states. The global relation is determined over the cell states in configuration $c^t$ updated at the next configuration $c^{t+1}$ simultaneously by a local function $\varphi$ as follows:

\begin{equation}
\varphi(x_{i-r}^t, \ldots, x_{i}^t, \ldots, x_{i+r}^t) \rightarrow x_i^{t+1}.
\label{localFunction}
\end{equation}

Wolfram represents one-dimensional CA with two parameters $(k,r)$, where $k = |\Sigma|$ is the number of states, and $r$ is the neighbourhood radius, hence ECA domain is defined by parameters $(2,1)$. There are $\Sigma^n$ different neighbourhoods (where $n=2r+1$) and $k^{k^n}$ distinct evolution rules. The evolutions in this paper have periodic boundary conditions.

%%%%%%%%%%%%%%%%%%%%%%
\subsection{Cellular automata with memory}
Conventional CA are ahistoric (memoryless): i.e., the new state of a cell depends on the neighbourhood configuration solely at the preceding time step of $\varphi$. CA with {\it memory} can be considered as an extension of the standard framework of CA where every cell $x_i$ is allowed to remember some period of its previous evolution. Basically memory is based on the state and history of the system, thus we design a memory function $\phi$, as follows:

\begin{equation}
\phi (x^{t-\tau}_{i}, \ldots, x^{t-1}_{i}, x^{t}_{i}) \rightarrow s_{i}
\end{equation}

\noindent such that $\tau < t$ determines the backwards degree of memory and each cell $s_{i} \in \Sigma$ is a function of the series of states in cell $x_i$ up to time-step $t-\tau$. Finally to execute the evolution we apply the original rule again as follows:

$$
\varphi(\ldots, s^{t}_{i-1}, s^{t}_{i}, s^{t}_{i+1}, \ldots) \rightarrow x^{t+1}_i.
$$

\begin{figure}[th]
\centerline{\includegraphics[width=4.85in]{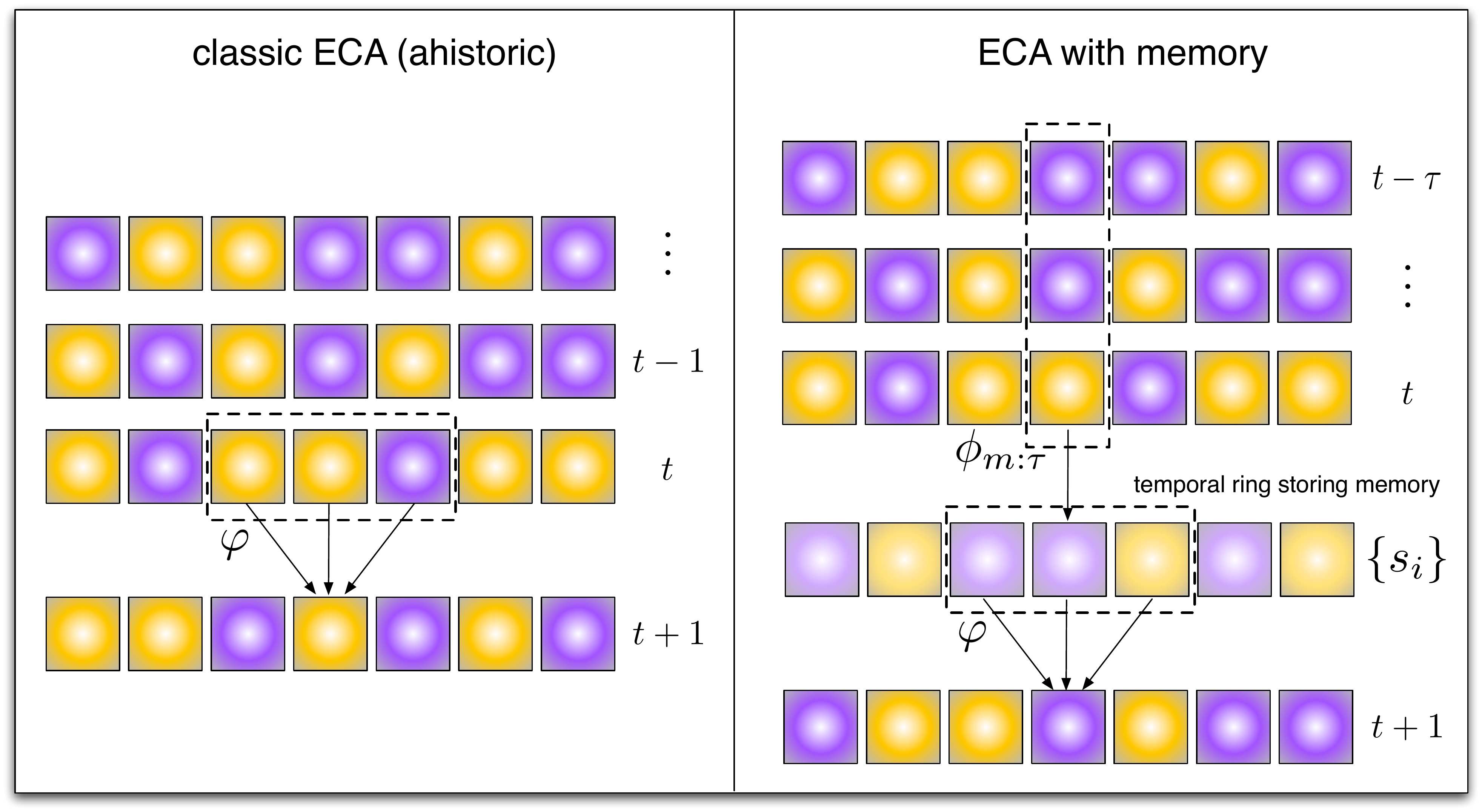}}
\caption{Cellular automata with memory in cells.}
\label{memEvol}
\end{figure}

In CA with memory,  while the mapping $\varphi$ remains unaltered, a historic memory of past iterations is retained by featuring each cell as a summary of its previous states; therefore cells {\it canalize} memory to the map $\varphi$. As an example, we can take the memory function $\phi$ as a {\it majority memory}:

\begin{equation}
\phi_{maj} \rightarrow s_{i}
\label{eq-majmem}
\end{equation}

\noindent where in case of a tie given by $\Sigma_1 = \Sigma_0$ in $\phi$, we shall take the last value $x_i$. So $\phi_{maj}$ represents the classic majority function for three variables \cite{kn:Mins67}, as follows:

\begin{equation*}
\phi_{maj} : (x_1 \wedge x_2) \vee (x_2 \wedge x_3) \vee (x_3 \wedge x_1) \rightarrow x
\end{equation*}

\noindent on cells $(x^{t-\tau}_{i}, \ldots, x^{t-1}_{i}, x^{t}_{i})$ and defines a temporal ring before calculating the next global configuration $c$. In case of a tie, it is allows to break it in favor of zero if $x_{\tau-1}=0$, or to one whether $x_{\tau-1}=1$.  The representation of a ECA with memory (given previously in \cite{kn:MAA09, kn:MAA09a}) is given as follows:

\begin{equation}
\phi_{CARm:\tau}
\end{equation}

\noindent where $CAR$ represents the decimal notation of a particular ECA and $m$ the kind of memory given with a specific value of $\tau$. Thus the majority memory ($maj$) working in ECA Rule 126 checking tree cells ($\tau=3$) of history is simply denoted as $\phi_{R126maj:3}$. 
Figure~\ref{memEvol} depicts in detail the memory working on ECA.

Note that memory is as simple as any CA local function but sometimes the global behaviour produced by the local rule is totally unpredictable.

%%%%%%%%%%%%%%%%%%%%%%
%\subsubsection{The effect of memory in CA literature}
\subsection*{Disclaimer}

The memory mechanism considered here differs from other CA with memory previously reported, often referred as {\it higher order} (in-time) CA.\footnote {See, for example, \cite{kn:Wolf84} p. 118; \cite{kn:Ilach01} p. 43; or class $\mathbb {MEMO}$ in \cite{kn:Ada94} p. 7.} These ones, in most cases, explicitly alter the  function $\phi$ and incorporate memory by direcly determining the new configuration in terms of the configurations at previous time-steps. Thus, in second order in time (memory of capacity two) rules, the transition rule operates as: $x^{t+1} = \Phi(c^{t}, c^{t-1})$. {\it Double} memory (in transition rule and in cells) can be implemented as: 
$x^{t+1} = \Phi(\{s^{t}\}, \{s^{t-1}\})$. Particularly interesting is the reversible formulation: $x^{t+1} = \varphi \ominus x^{t-1}$; reversible CA with memory are studied in \cite{kn:Alo09}.

Some authors define rules with {\it memory} as those with dependence in $\varphi$ on the state of the cell to be updated \cite{kn:Wolf00, kn:Kau84}. So one-dimensional rules with no {\it memory} adopt the form: $x^{t+1} = \varphi(x_{i-1}^{t},x_{i+1}^{t})$. Memory is not here indentified with {\it delay}, i.e., refering cells exclusively to their  \index{Delay} state values a number of time-steps in the past, \cite{kn:RJ09}. So, for example, the cell to be updated may be referenced not at $t$ but at $t-1$: $x^{t+1} = \varphi(x_{i-1}^{t}, x_{i}^{t}, x_{i+1}^{t})$ \cite{kn:Let07}. Again, the mapping function is not $extended$, for example, to consider the influence of cell $i$ at time $t-1$: $x^{t+1} = \psi(x_{i}^{t-1}, x_{i-1}^{t}, x_{i}^{t}, x_{i+1}^{t})$ as done in \cite{kn:XQM05}.

The use of the locution {\it associative} memory usually refers, when used in the CA context, to the study of configuration attractors \cite{kn:GMD02, kn:MC08}, which are argued by Wuensche \cite{kn:Wue94b} to constitute the network's global states contents addressable memory in the sense of Hopfield \cite{kn:Hop82}.

Finally, it is not intended here to emulate {\it human} memory, i.e., the associative, pattern matching, highly parallel function of human memory.
The aim is just to {\it store} the past, or just a part of it, to make it {\it work} in dynamics. Thus, {\it working storage} might replace here the use of the term {\it memory}, avoiding the anthropomorphic, and rather unavoidable, connotations of the word memory \cite{kn:Alo09}.

%%%%%%%%%%%%%%%%%%%%%%
\section{The basic function: ECA Rule 126}

The local-state transition function $\varphi$ corresponding to Rule 126 is represented as follows:

\[
\varphi_{R126} = \left\{
	\begin{array}{lcl}
		1 & \mbox{if} & 110, 101, 100, 011, 010, 001 \\
		0 & \mbox{if} & 111, 000
	\end{array} \right. .
\]

\begin{figure}[th]
\begin{center}
\subfigure[]{\scalebox{0.38}{\includegraphics{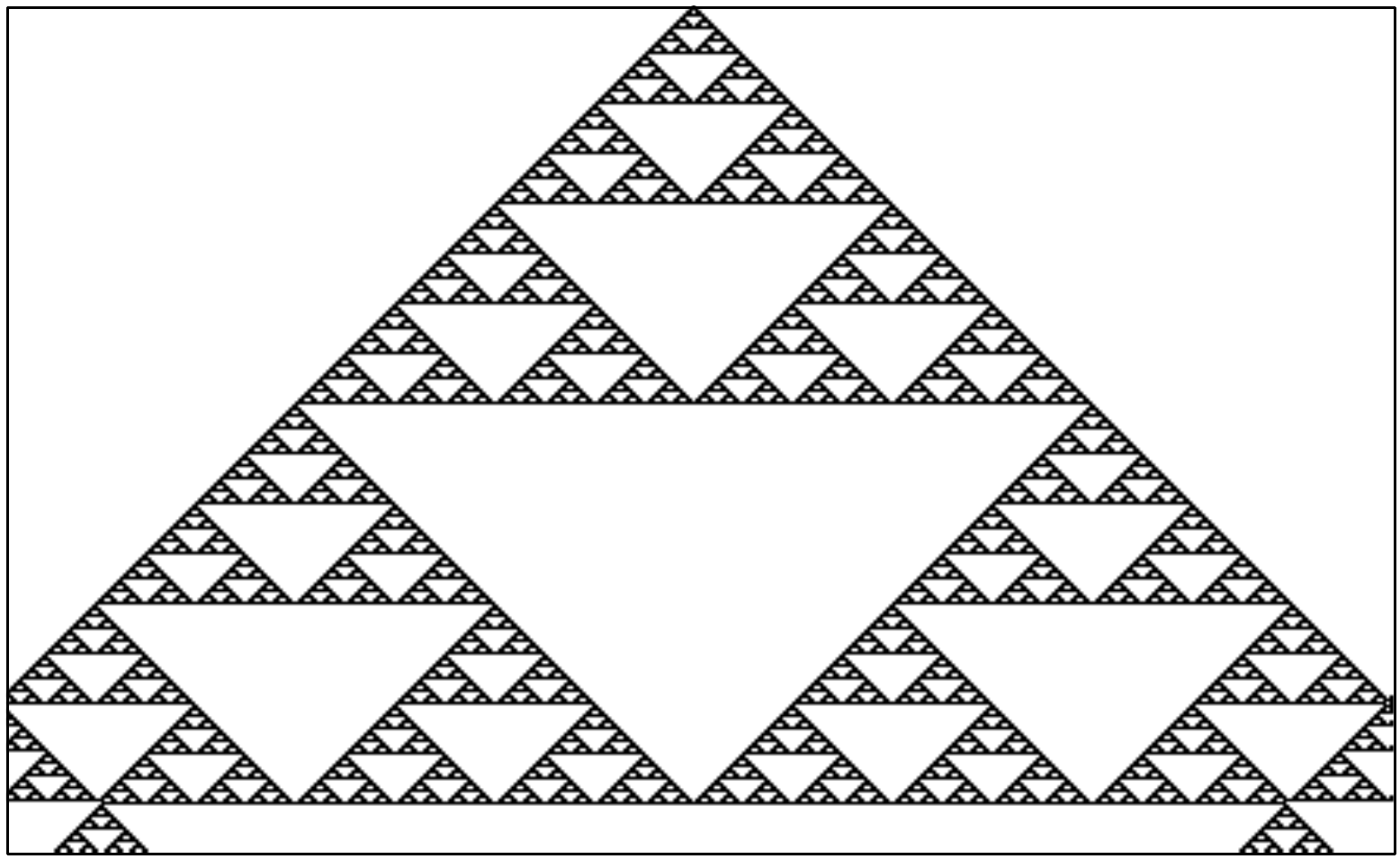}}} %\hspace{0.1cm}
\subfigure[]{\scalebox{0.38}{\includegraphics{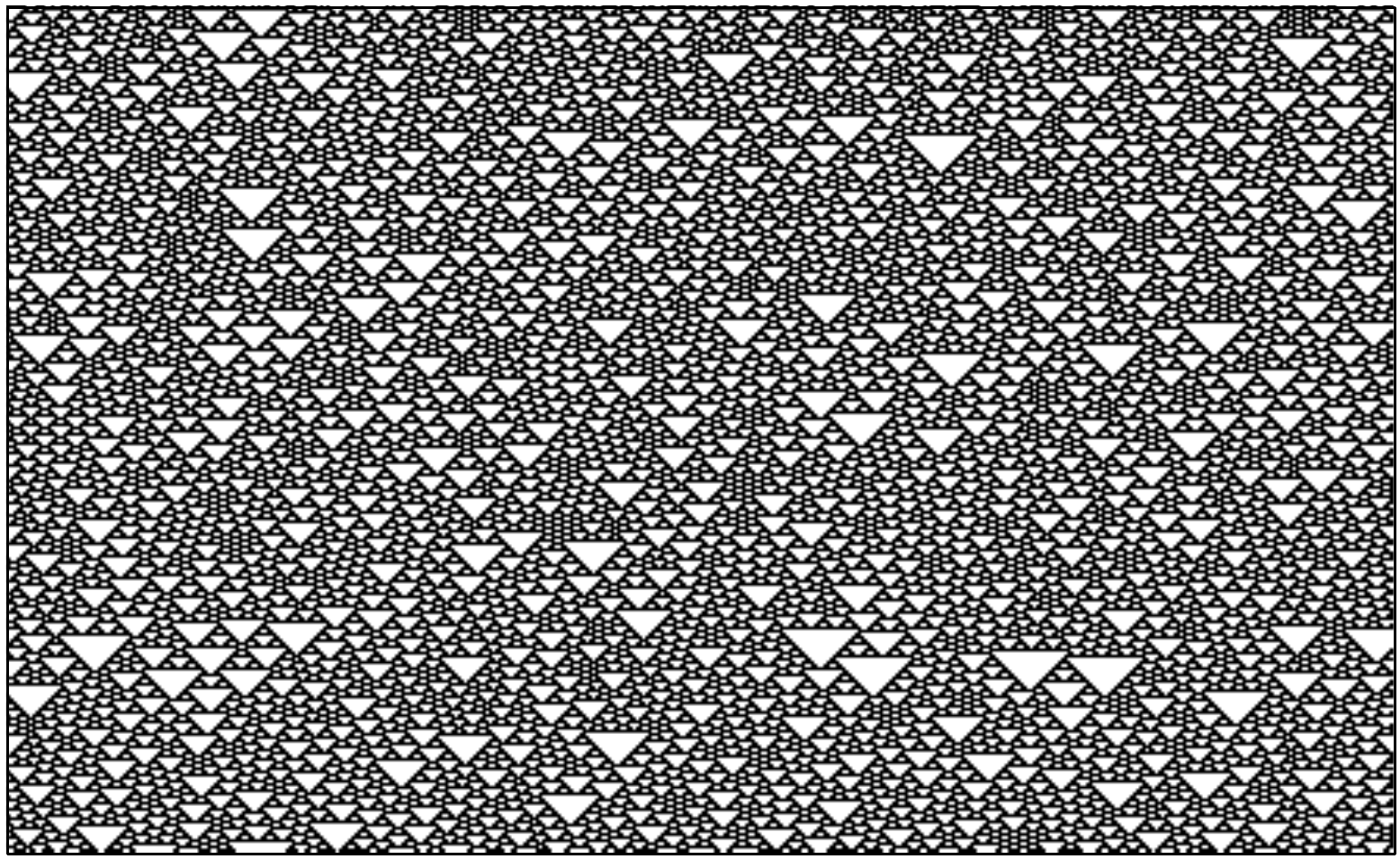}}} \end{center}
\caption{Typical fractal~(a) and chaotic~(b) global evolution of ECA Rule 126. (a)~initially all cells in `0' but one in state `1,' (b)~evolution from random initial configuration with 50\% of `0` and `1' states.}
\label{randomEvol}
\end{figure}

Rule 126 has a chaotic global behaviour typical from Class III in Wolfram's classification~\cite{kn:Wolf94}. In $\varphi_{R126}$ we can easily recognize an initial high probability of alive cells, i.e. cells in state `1'; with a 75\% to appear in the next time and, complement of only 25\% to get state 0. It will be always a new alive cell iff $\varphi_{R126}$ has one or two alive cells such that the equilibrium comes when there is an overpopulation condition. Figure~\ref{randomEvol} shows these cases in typical evolutions of Rule 126, both evolving from a single cell in state `1' (fig.~\ref{randomEvol}a) and from a random initial configuration (fig.~\ref{randomEvol}b) where a high density of 1's is evidently in the evolution.

While looking on chaotic space-time configuration in fig.~\ref{randomEvol} we understand the difficulty for analysing the rule's behaviour and selecting any coherent activity among periodic structures without special tools.

%%%%%%%%%%%%%%%%%%%%%%
\subsection{Mean field approximation in ECA Rule 126}

This section presents a probabilistic analysis with mean field theory, in order to search basic properties about $\varphi_{R126}$ evolution space and its related chaotic behaviour. Such analysis will offer a better spectrum where we can start to explore the evolution space from more useful and specific initial conditions where some interesting behaviours may emerge.

Mean field theory is a well-known technique for discovering statistical properties of CA without analysing evolution spaces of individual rules~\cite{kn:Mc09}. The method assumes that states in $\Sigma$ are independent and do not correlate each other in the local function $\varphi_{R126}$. Thus we can study probabilities of states in a neighbourhood in terms  of the probability of a single state (the state in which the neighbourhood evolves), and probability of the neighbourhood is product of the probabilities of each cell in it.

In this way, \cite{kn:Mc90} presents an explanation of Wolfram's classes by a mixture of probability theory and de Bruijn diagrams, resulting a classification based on mean field theory curve:

\begin{itemize}
\item class I: monotonic, entirely on one side of diagonal;
\item class II: horizontal tangency, never reaches diagonal;
\item class IV: horizontal plus diagonal tangency, no crossing;
\item class III: no tangencies, curve crosses diagonal.
\end{itemize}

For the one-dimensional case, all  neighbourhoods are considered as follows:

\begin{equation}
p_{t+1}=\sum_{j=0}^{k^{2r+1}-1}\varphi_{j}(X)p_{t}^{v}(1-p_{t})^{n-v}
\label{MFp1D}
\end{equation}

\noindent such that $j$ is an index number relating each neighbourhood and $X$ are cells $x_{i-r}, \ldots, x_{i}, \ldots, x_{i+r}$. Thus $n$ is the number of cells into every neighbourhood, $v$ indicates how often state `1' occurs in $X$, $n-v$ shows how often state `0' occurs in the neighbourhood $X$, $p_{t}$ is the probability of cell being in state `1' while $q_{t}$ is the probability of cell being in state `0' i.e., $q=1-p$. The polynomial for Rule 126 is defined as follows:

\begin{equation}
p_{t+1}=3p_{t}q_{t}.
\label{pR126}
\end{equation}

Because $\varphi_{R126}$ is classified as a chaotic rule, we expect no tangencies and its curve must cross the identity; remembering that $\varphi_{R126}$ has a 75\% of probability to produce a state one. 

\begin{figure}[th]
\centerline{\includegraphics[width=3in]{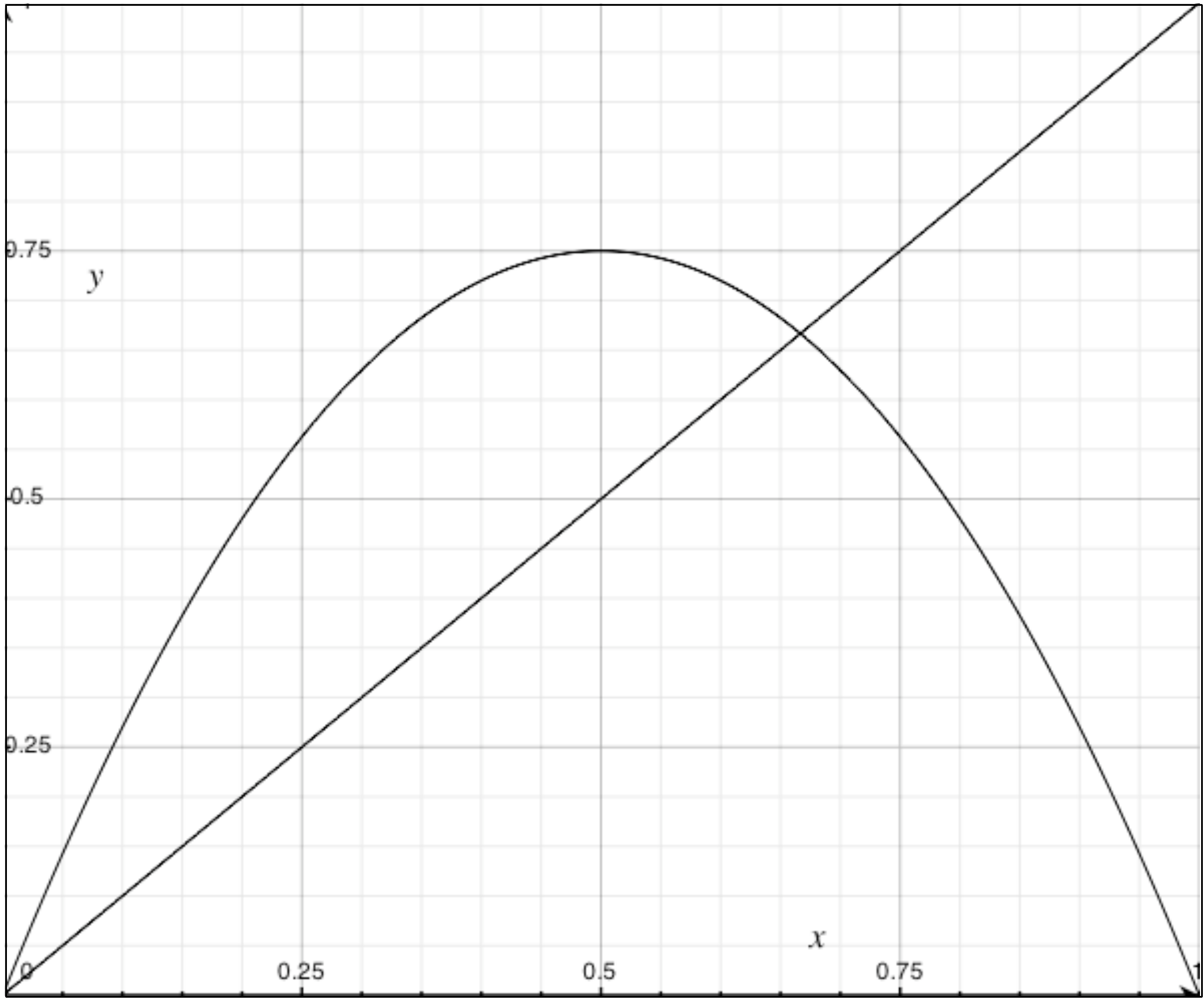}}
\caption{Mean field curve for ECA Rule 126. }
\label{meanField}
\end{figure}

Mean field curve (fig.~\ref{meanField}) confirms that probability of state `1' in space-time configurations of $\varphi_{R126}$ is 0.75 for high densities related to big populations of 1's. The curve demonstrates also that $\varphi_{R126}$ is chaotic because the curve cross the identity with a first fixed point at the origin $f=0$ and the non existence of any unstable fixed point inducing non stable regions in the evolution. Nevertheless, the stable fixed point is $f=0.6683$, which represents a `concentration' of `1's  diminishing during the automaton evolution.

So the initial inspection indicates no evidence of complex behaviour emerging in $\varphi_{R126}$. Of course a deeper analysis is necessary for obtaining more features from a chaotic rule, so the next sections explain other techniques to study in particular periodic structures.

%%%%%%%%%%%%%%%%%%%%%%
\subsection{Basins of attraction}
A basin (of attraction) field of a finite CA is the set of basins of attraction into which all possible states and trajectories will be organized by the local function $\varphi$. The topology of a single basin of attraction may be represented by a diagram, the {\it state transition graph}. Thus the set of graphs composing the field specifies the global behaviour of the system \cite{kn:WL92}.

Generally a basin can also recognize CA with chaotic or complex behaviour following previous results on attractors \cite{kn:WL92}. Thus we have that Wolfram's classes can be represented as a basin classification:

\begin{itemize}
\item class I: very short transients, mainly point attractors (but possibly also periodic attractors) very high in-degree, very high leaf density (very ordered dynamics);
\item class II: very short transients, mainly short periodic attractors (but also point attractors), high in-degree, very high leaf density;
\item class IV: moderate transients, moderate-length periodic attractors, moderate in-degree, very moderate leaf density (possibly complex dynamics);
\item class III: very long transients, very long periodic attractors, low in-degree, low leaf density (chaotic dynamics).
\end{itemize}

\begin{figure}[th]
\begin{center}
\subfigure[]{\scalebox{0.32}{\includegraphics{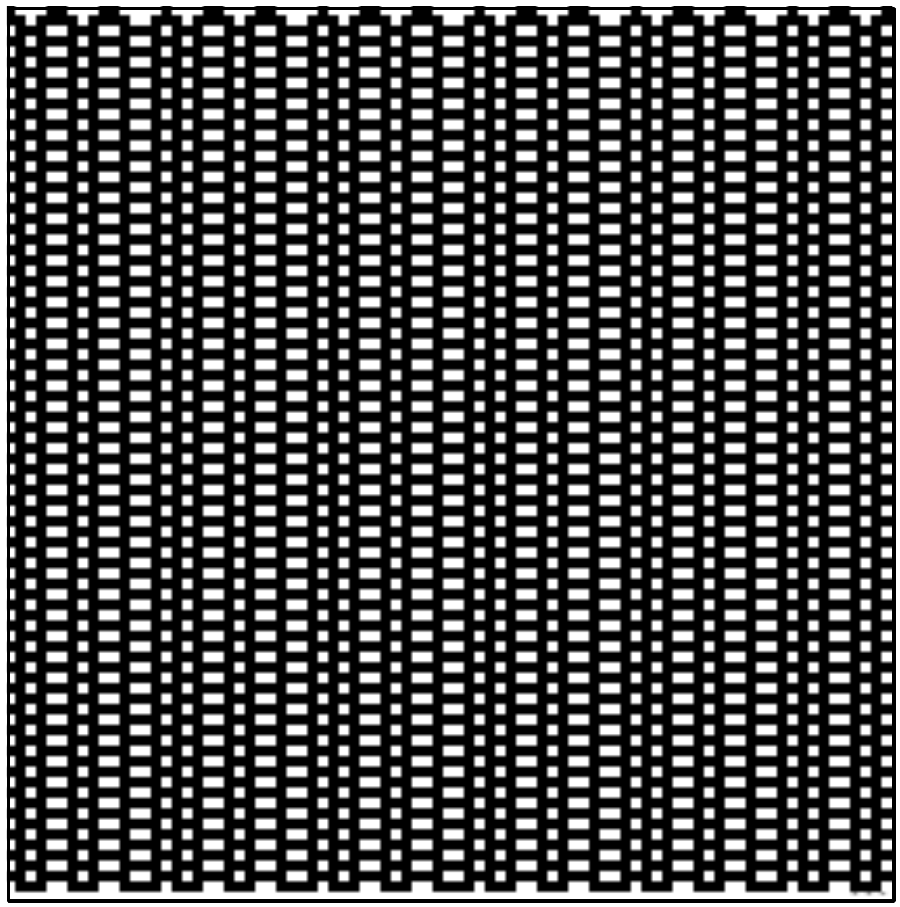}}} %\hspace{0.1cm}
\subfigure[]{\scalebox{0.32}{\includegraphics{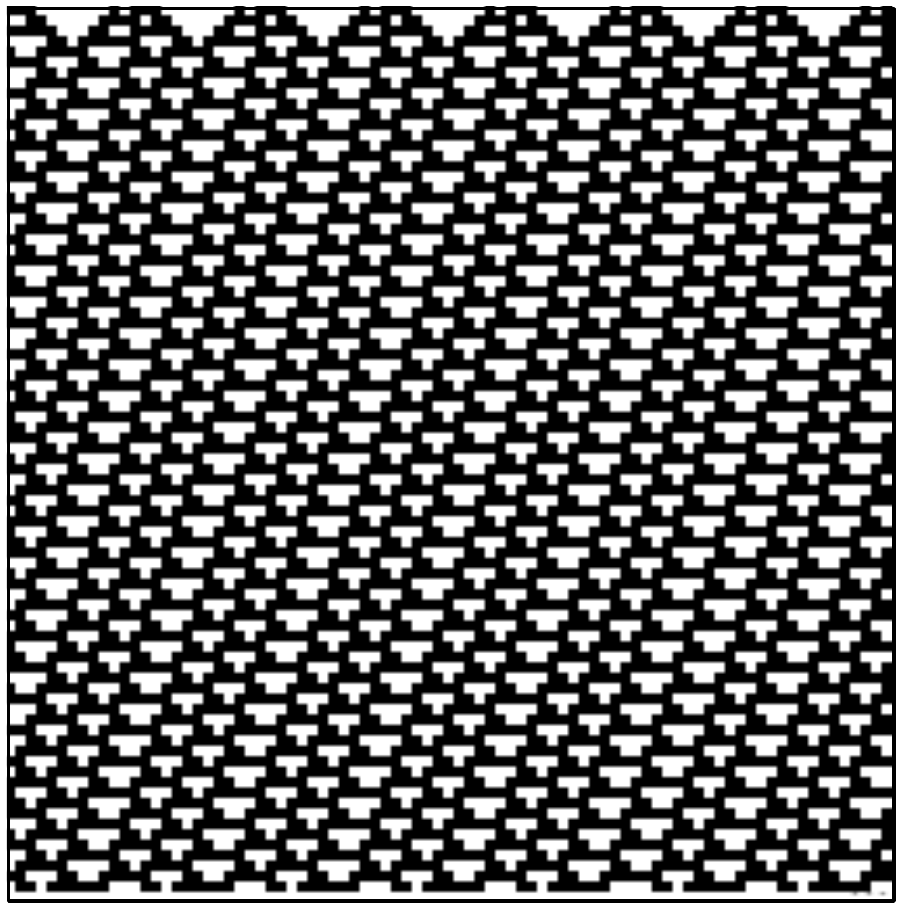}}}
\subfigure[]{\scalebox{0.32}{\includegraphics{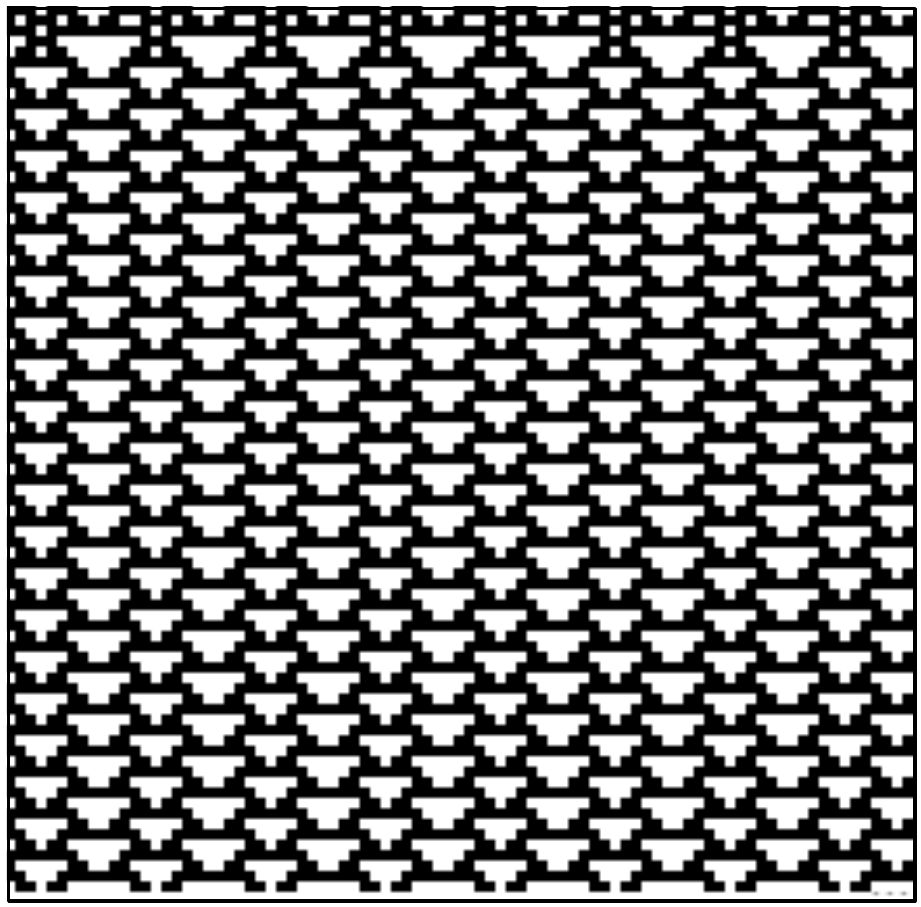}}}
\subfigure[]{\scalebox{0.32}{\includegraphics{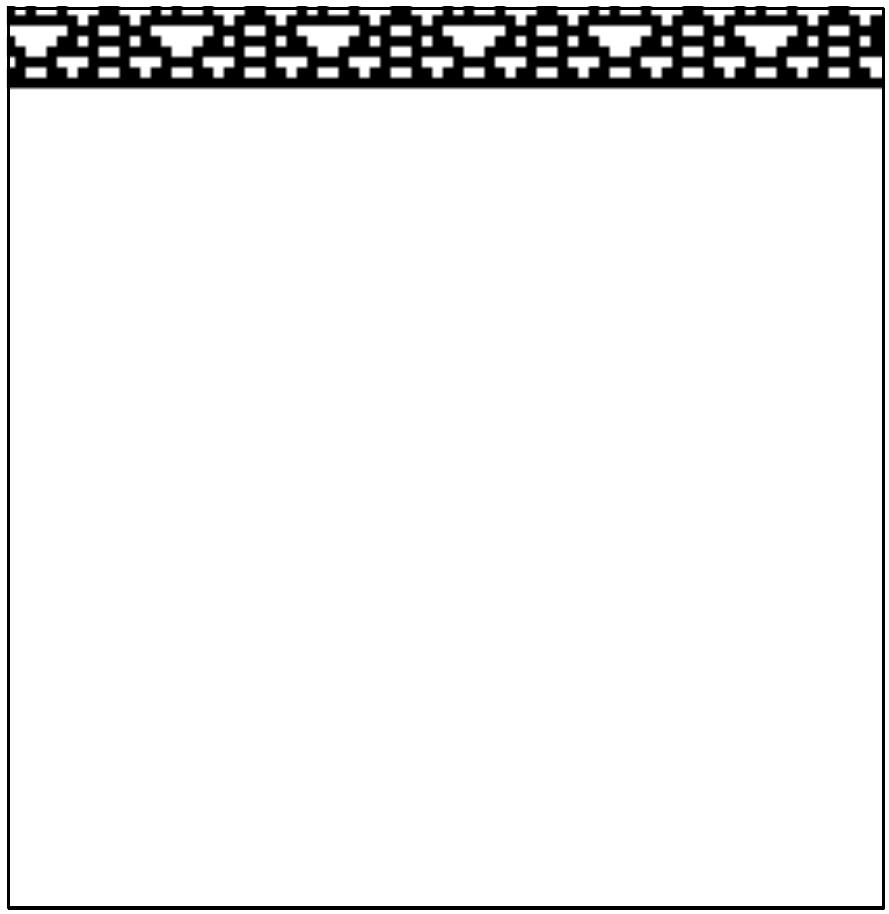}}}
\end{center}
\caption{Periodic patterns from some basin attractors.}
\label{cyclesR126}
\end{figure}

\begin{figure}%[th]
\centerline{\includegraphics[width=5.2in]{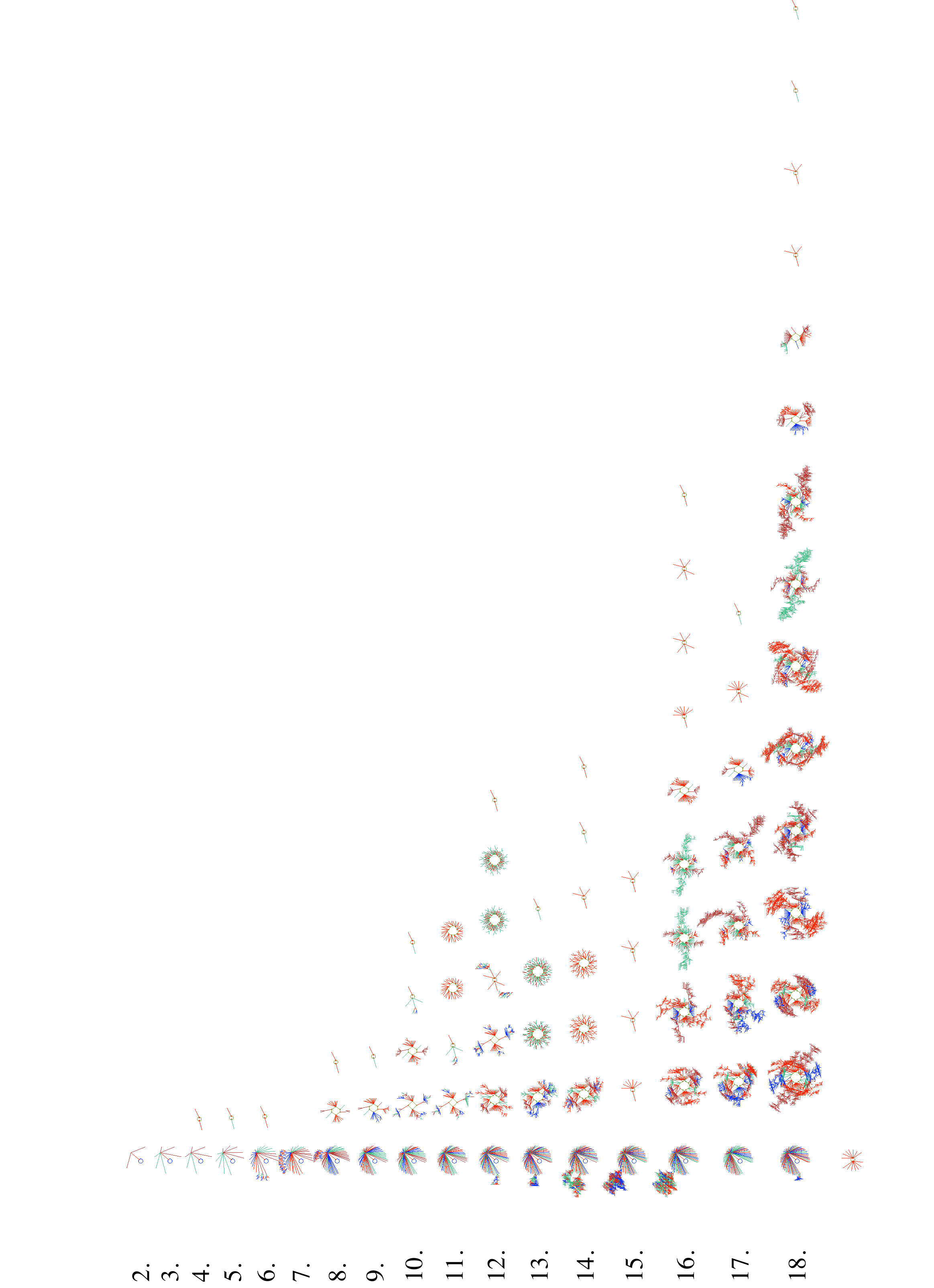}}
\caption{The whole set of non-equivalent basins in ECA Rule 126 from $l=2$ to $l=18$.}
\label{r126_2-18}
\end{figure}

The basins depicted in fig.~\ref{r126_2-18} show the whole set of non-equivalent basins in Rule 126 from $l=2$ to $l=18$ ($l$ means length of array) attractors, all they display not high densities from an attractor of mass one and attractors of mass 14.\footnote{Basins and attractors were calculated with {\it Discrete Dynamical System} DDLab available from \url{http://www.ddlab.org/}} This way Rule 126 displays some non symmetric basins and some of them have long transients that induce a relation with chaotic rules.

Particularly we can see specific cycles in fig.~\ref{cyclesR126} where it is possible to find:

\begin{itemize}
\item[(a)] static configurations as still life patterns ($l=8$);
\item[(b)] traveling configurations as gliders ($l=15$);
\item[(c)] meshes ($l=12$);
\item[(d)] or empty universes ($l=14$).
\end{itemize}

The cycle diagrams expose only displacements to the left, and this empty universe evolving to the stable state 0 is constructed all times on the first basin for each cycle, see fig.~\ref{r126_2-18}.

This way some cycles could induce some non trivial activity in Rule 126, but the associated initial conditions are not generally predominant. However some information is useful indeed looking periodic patterns that have a high frequency inside this evolution space and hence for recognizing a kind of filter useful to get a better view of a possible complex activity in Rule 126.

%%%%%%%%%%%%%%%%%%%%%%
\subsection{De Bruijn diagrams}

De Bruijn diagrams \cite{kn:Mc09,kn:Voor96} are very adequate for describing evolution rules in one-dimensional CA, although originally they were used in shift-register theory (the treatment of sequences where their elements overlap each other). Paths in a de Bruijn diagram may represent chains, configurations or classes of configurations in the evolution space. 

For an one-dimensional CA of order $(k,r)$, the de Bruijn diagram is defined as a directed graph with $k^{2r}$ vertices and $k^{2r+1}$ edges. The vertices are labeled with the elements of the alphabet of length $2r$. An edge is directed from vertex $i$ to vertex $j$, if and only if, the $2r-1$ final symbols of $i$ are the same that the $2r-1$ initial ones in $j$ forming a neighbourhood of $2r+1$ states represented by $i \diamond j$. In this case, the edge connecting $i$ to $j$ is labeled with $\varphi(i \diamond j)$ (the value of the neighbourhood defined by the local function) \cite{kn:Voor06}.

\begin{figure}[th]
\centerline{\includegraphics[width=1.6in]{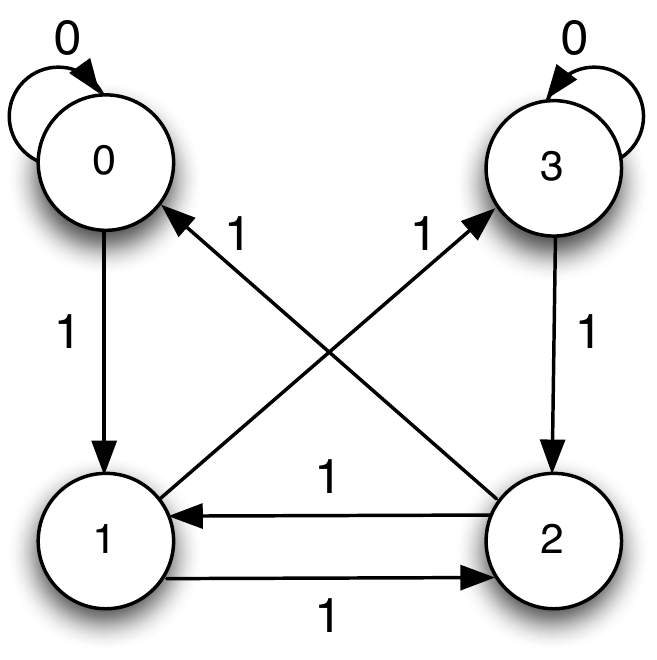}}
\caption{De Bruijn diagram for the ECA Rule 126.}
\label{dB126}
\end{figure}

The de Bruijn diagram associated to Rule 126 is depicted in fig. \ref{dB126}.\footnote{De Bruijn and subset diagrams were calculated using NXLCAU21 designed by Harold V. McIntosh; available from \url{http://delta.cs.cinvestav.mx/~mcintosh}}

\begin{figure}[th]
\begin{center}
\subfigure[]{\scalebox{0.42}{\includegraphics{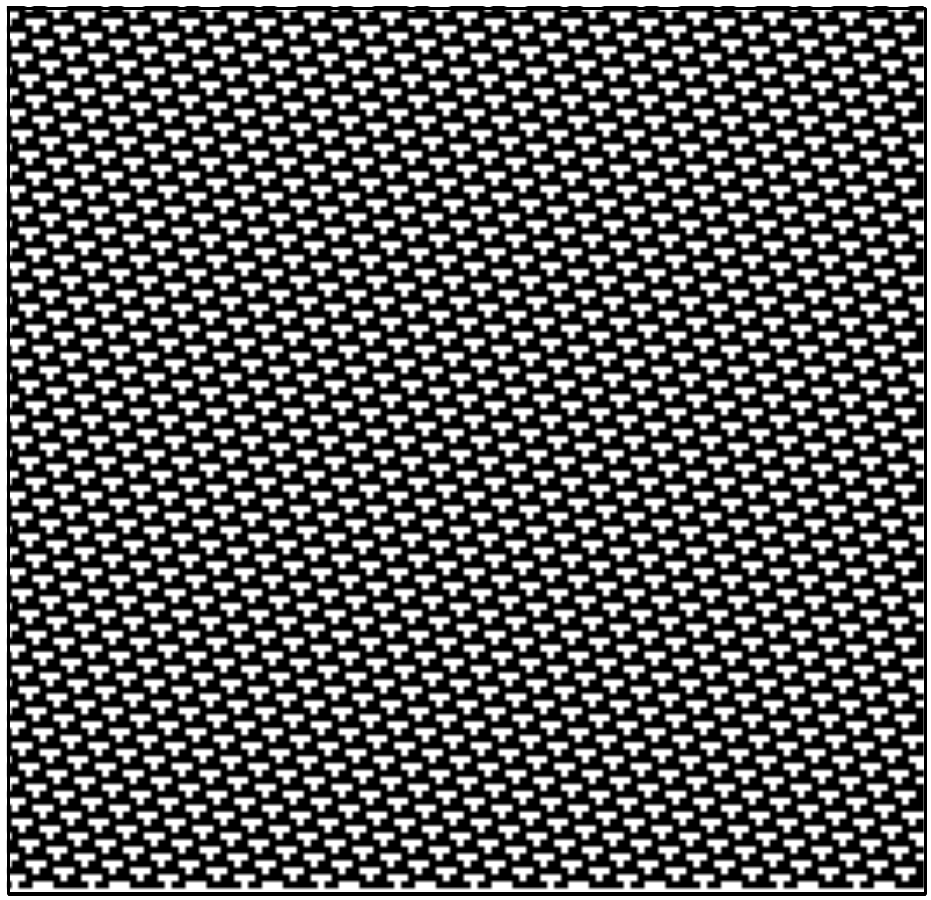}}} %\hspace{0.1cm}
\subfigure[]{\scalebox{0.42}{\includegraphics{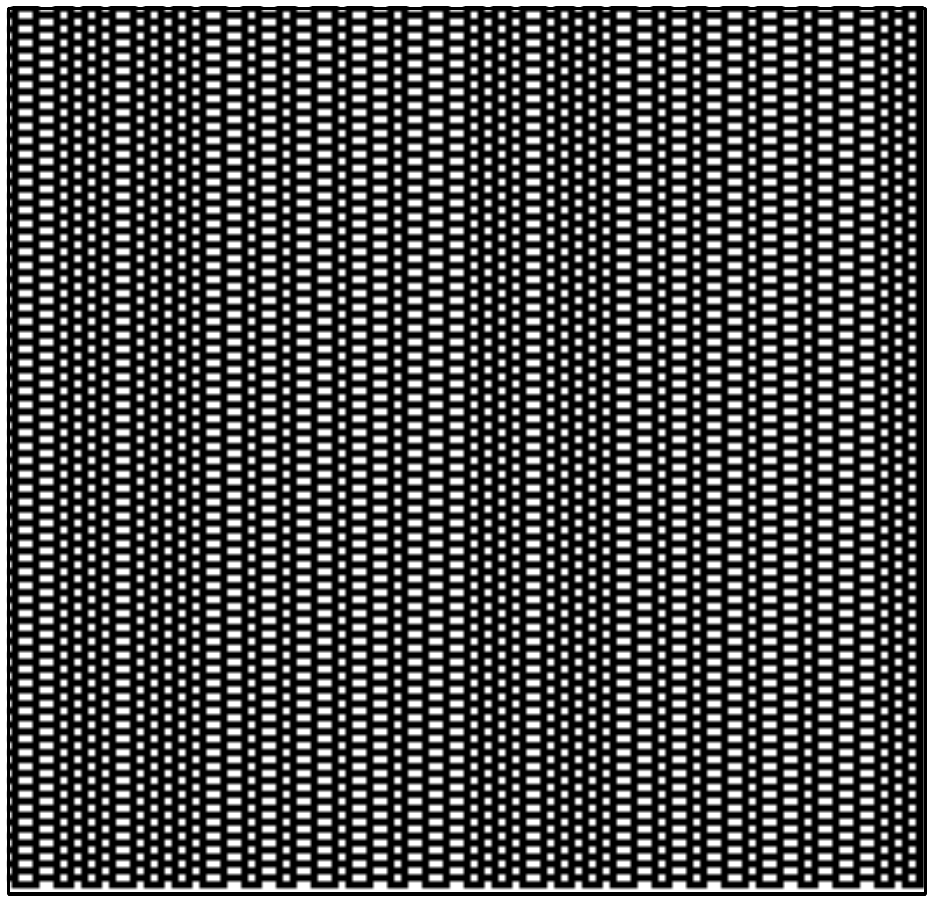}}} %\hspace{0.1cm}
\subfigure[]{\scalebox{0.42}{\includegraphics{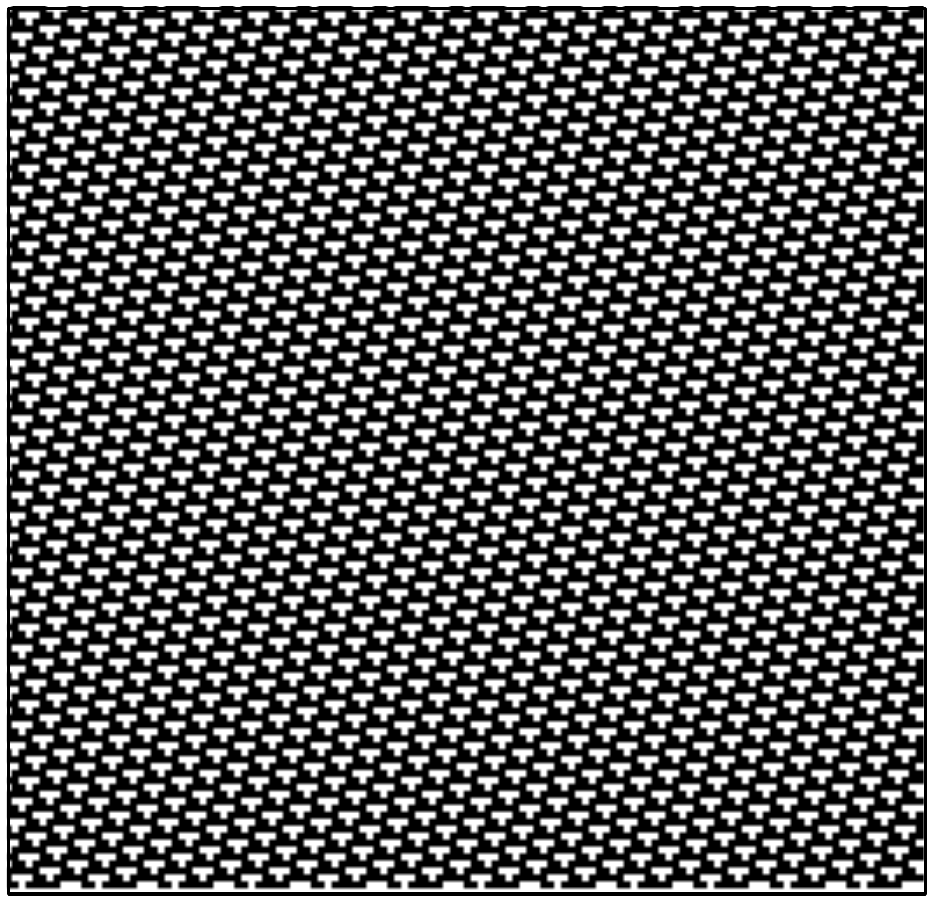}}}
\end{center}
\caption{Patterns calculated from extended de Bruijn diagrams, in particular from cycles of order $(x,2)$ (that means a $x$-shift  in $y$-generations).}
\label{dB126-1}
\end{figure}

Figure~\ref{dB126} exposes that there are two neighbourhoods evolving into $0$ and six neighbourhoods into $1$; so the higher frequency is for state 1; indicating the possibility of having an injective automaton; that is, the existence of  {\it Garden of Eden} configurations \cite{kn:Mc09, kn:Voor96}. Classical analysis in graph theory has been applied over de Bruijn diagrams for studying topics such as reversibility \cite{kn:Seck05}; in other sense, cycles in the diagram indicate periodic constructions in the evolution of the automaton if the label of the cycle agrees with the sequence defined by its nodes, taking periodic boundary conditions. Let us take the equivalent construction of a de Bruijn diagram in order to describe the evolution in two steps of Rule 126 (having now nodes composed by sequences of four symbols); the cycles of this new diagram are presented in fig.~\ref{dB126-1}.

The extended de Bruijn diagrams \cite{kn:Mc09} are useful for calculating all periodic sequences by the cycles defined in the diagram. These ones also show the {\it shift} of a sequence for a certain number of {\it generations}. Thus we can get de Bruijn diagrams describing periodic sequences for Rule 126.

% 1000 - pure s1
% 10000 - s1 + b1
% 11000 - s1 + b1
% 110000 - filter b1

Cycles inside de Bruijn diagrams can be used for obtaining regular expressions representing a periodic pattern. Figure~\ref{dB126-1} displays three patterns calculated as: (a) shift $-3$ in 2 generations representing a pattern with displacement to the left, (b) shift 0 in 2 generations describing a static pattern traveling without displacement, and (c) shift $+3$ in 2 generations is exactly the symmetric pattern given in the first evolution.

So we can also see in fig.~\ref{dB126-1} that it is possible to find patterns traveling in both directions, as gliders or mobile structures. But generally these constructions (strings) cannot live in combination with others structures and therefore it is really hard to have this kind of objects with such characteristics. Although, moreover Rule 126 has at least one glider! This will be explained in the next sections.

%%%%%%%%%%%%%%%%%%%%%%
\subsection{Filters for recognizing dynamics in Rule 126}

Filters are a useful tool for discovering hidden order in chaotic or complex rules. Filters were introduced in CA studies by Wuensche who employed them to automatically classify cell-state transition functions, see~\cite{kn:Wue99}. Also filters related to tiles were successfuly applied and deduced in analysing space-time behaviour of ECA governed by Rules 110 and 54~\cite{kn:MMS06,kn:MAM06,kn:MAM08}.

This way, we have found that Rule 126 has two types of two-dimensional tiles (which together work as filters over $\varphi_{R126}$):

\begin{itemize} 
\item the tile {\footnotesize $t_1 = \begin{bmatrix} 1111 \\ 1001 \end{bmatrix}$}, and
\item the tile {\footnotesize $t_2 = \begin{bmatrix} 0000000 \\ 0111110  \\ 1100011 \\ 0110110 \\ 1111111 \end{bmatrix}$}.
\end{itemize}

Filter $t_1$ works more significantly on configurations generated by $\varphi_{R126}$, the second one is not frequently found although it is exploited when Rule 126 is altered with memory (as we can see in following sections).

\begin{figure}[th]
\centerline{\includegraphics[width=4in]{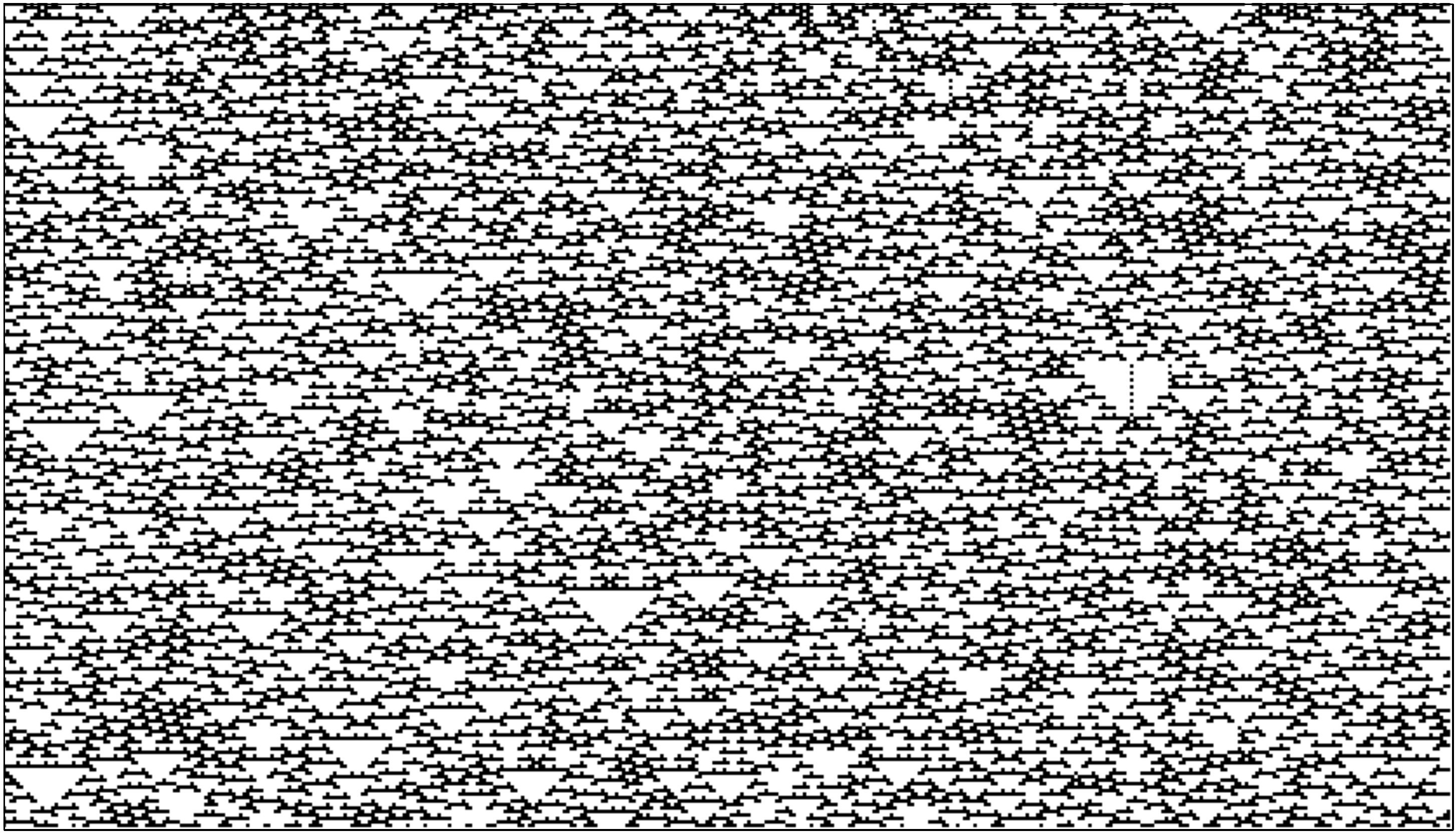}}
\caption{Filtered space-time configuration in $\varphi_{R126}$.}
\label{R126filtered}
\end{figure}

The application of the first filter is effective to discover gaps with little patterns traveling on triangles of `1' states in the evolution space. Although even in this case it may be unclear how a dynamics would be interpreted, a careful inspection on the evolution brings to light very small gliders (as still life), as shown in fig.~\ref{R126filtered}.

This glider emerging in Rule 126 and localized by a filter is precisely the periodic pattern calculated with the basin (fig.~\ref{cyclesR126}a) and the de Bruijn diagram (fig.~\ref{dB126-1}b); in particular the last one offers more information because such cycles allow to classify the whole phases when this glider is coded in the initial condition. The next sections demonstrate the effect of filters for recognizing an amazing universe evolving in this CA with memory.

%%%%%%%%%%%%%%%%%%%%%%
\section{CA $\phi_{R126m:4}$ and complex dynamics}
This section discusses both relevant aspects of classic Rule 126 ($\varphi_{R126}$) and Rule 126 with memory ($\phi_{R126m:\tau}$).

%%%%%%%%%%%%%%%%%%%%%%
\subsection{Dynamics emerging with majority memory}
As it was explained in \cite{kn:Alo09b, kn:MAA09} a new family of evolution rules derived from classic ECA can be found selecting a kind of memory.

\begin{figure}[th]
\centerline{\includegraphics[width=4.78in]{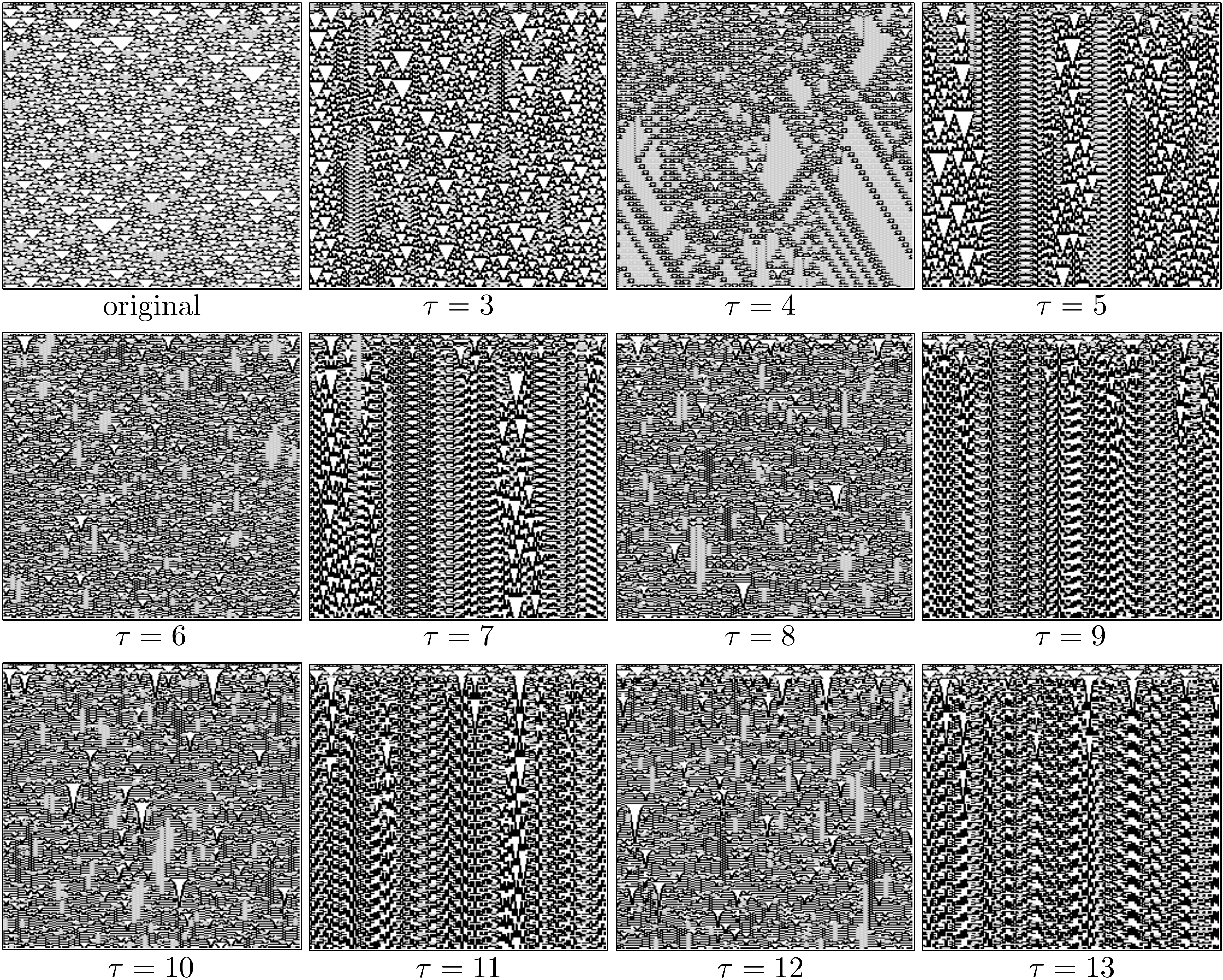}}
\caption{CA with majority memory $\phi_{R126maj:\tau}$ where 13 values of $\tau$ are evolved and filtered over a ring of 246 cells for 236 generations.}
\label{majMemory}
\end{figure}

Figure~\ref{majMemory} illustrates dynamics for some values of $\tau$ in $\phi_{R126maj}$.\footnote{Evolutions of $\phi_{R126maj:\tau}$ were calculated with {\it OSXLCAU21 system} available in \url{http://uncomp.uwe.ac.uk/genaro/OSXCASystems.html}} The space-time configurations are also filtered to show a raw dynamics. We found that large odd values of $\tau$ tend to define {\it macrocells}-like patterns~\cite{kn:Wolf94, kn:Mc09}. Even values of $\tau$ are responsible of a mixture of periodic and chaotic dynamics.

\begin{figure}%[th]
\centerline{\includegraphics[width=3.5in]{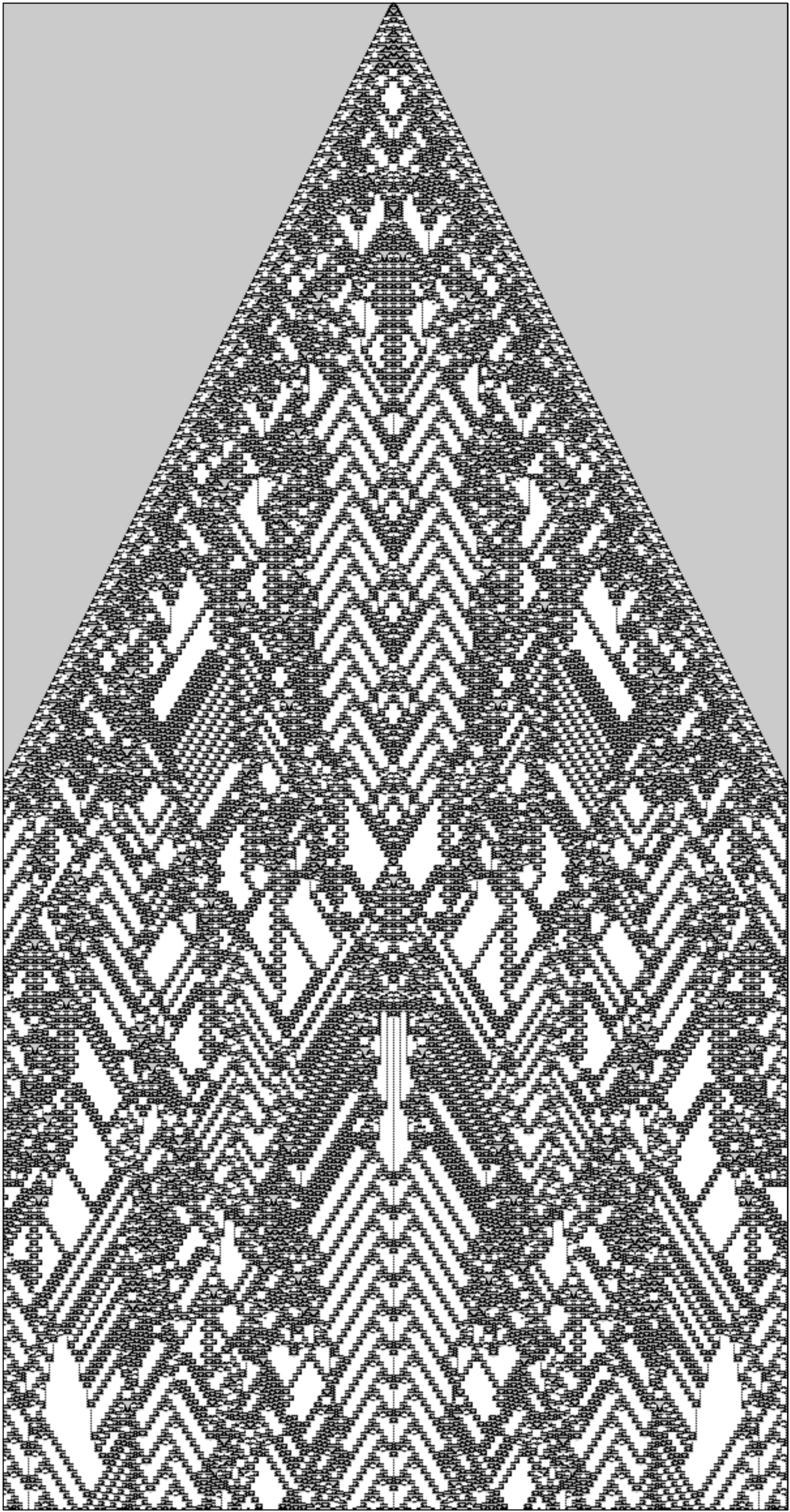}}
\caption{Filtered space-time configuration of $\phi_{R126maj:4}$ with 602 cells, periodic boundaries, starting from one non-quiescent cell and running for 1,156 steps.}
\label{oneCell}
\end{figure}

On exploring systematically distinct values of $\tau$, we found that $\phi_{R126maj:4}$ produces an impressive and non-trivial emergence of patterns traveling and colliding; so yet when the memory is working as a new function, it is possible to recognize some fragments inherited of the original rule $\varphi_{R126}$.

We start our simulations with a single non-quiescent cell, an example of this space-time configuration is provided in fig.~\ref{oneCell} showing the first 1,156 steps, where in this case the automaton needed other 12,000 steps to reach a stationary configuration. Filter is convenient to eliminate the non relevant information about gliders. In fig.~\ref{oneCell} we can see a number of gliders, glider guns, still-life configurations, and a wide number of combinations of such patterns colliding and traveling with different velocities and densities. Consequently we can classify a number of periodic structures, objects and interesting reactions.

\begin{figure}[th]
\centerline{\includegraphics[width=3.8in]{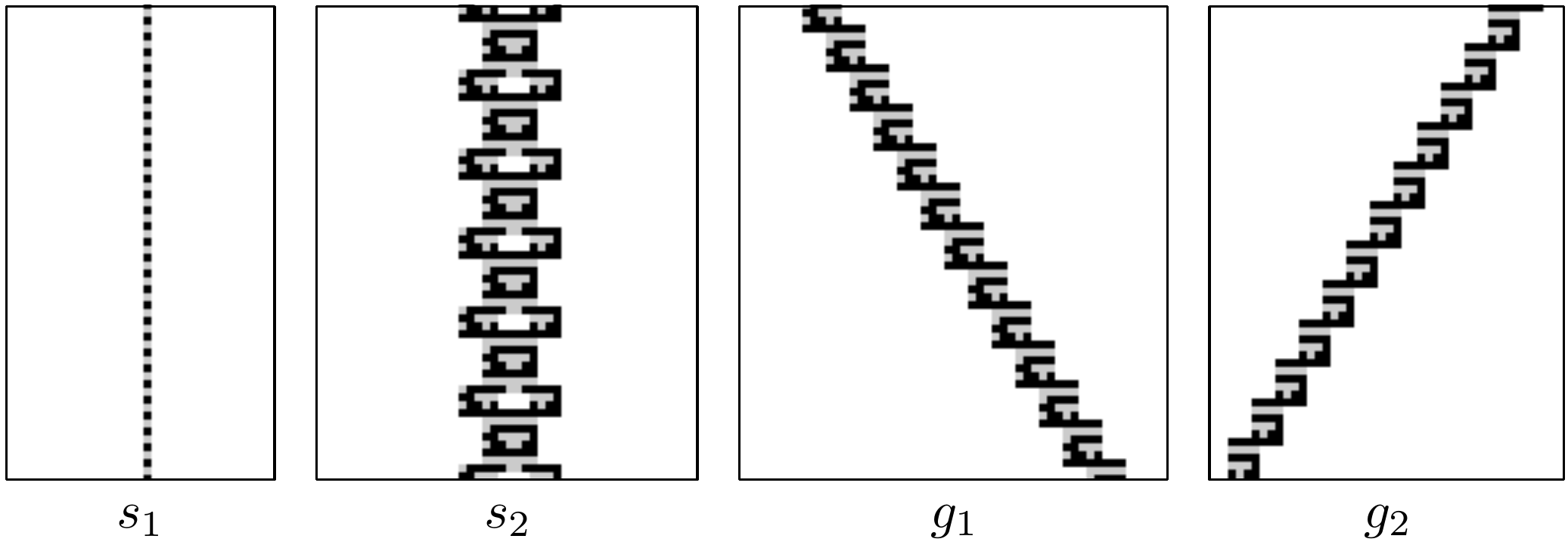}}
\caption{Basic gliders in $\phi_{R126maj:4}$. Two stationary configurations (as still life patterns) $s_1$ and $s_2$ respectively, and two gliders $g_1$ and $g_2$.}
\label{basicGliders}
\end{figure}

Basic primitive gliders are displayed in fig.~\ref{basicGliders}, there are still-life patterns $s_1$ and $s_2$, and gliders $g_1$ and $g_2$ respectively. These structures can be ordered in a set $\cal{G}$$_{\phi_{R126maj:4}}=\{s_1,s_2,g_1,g_2\}$.

\begin{table}[th]
\centering
\small
\begin{tabular}{|c|c|c|c|}
\hline
structure & $v_{g}$ & lineal volume & mass \\
\hline \hline
$s_1$ & $0/2$ = 0 & 1 & 1 \\
\hline
$s_2$ & $0/10$ = 0 & 12 & 28 \\
\hline
$g_{1}$ & $3/5 \approx 0.6$ & 8 & 17 \\
\hline
$g_{2}$ & $-3/5 \approx -0.6$ & 8 & 17 \\
\hline
$\mbox{gun}_{1}$ & $0/19 = 0$ & 6 & - \\
\hline
$\mbox{gun}_{2}$ & $0/27 = 0$ & 6 & - \\
\hline
$\mbox{gun}_{3}$ & $0/110 = 0$ & 10 & - \\
\hline
$\mbox{gun}_{4}$ & $0/84 = 0$ & 15 & - \\
\hline
\end{tabular}
\caption{Properties of gliders $\cal G$$_{\phi_{R126maj:4}}$. Velocity $v_g$ is calculated as displacement between period. The linear volume is the maximum distance between two cells in state `1' (diameter of the set of cells in state `1'), finally mass is the number of cells in state `1.'}
\label{tableGliders}
\end{table}

\begin{figure}[th]
\centerline{\includegraphics[width=3.3in]{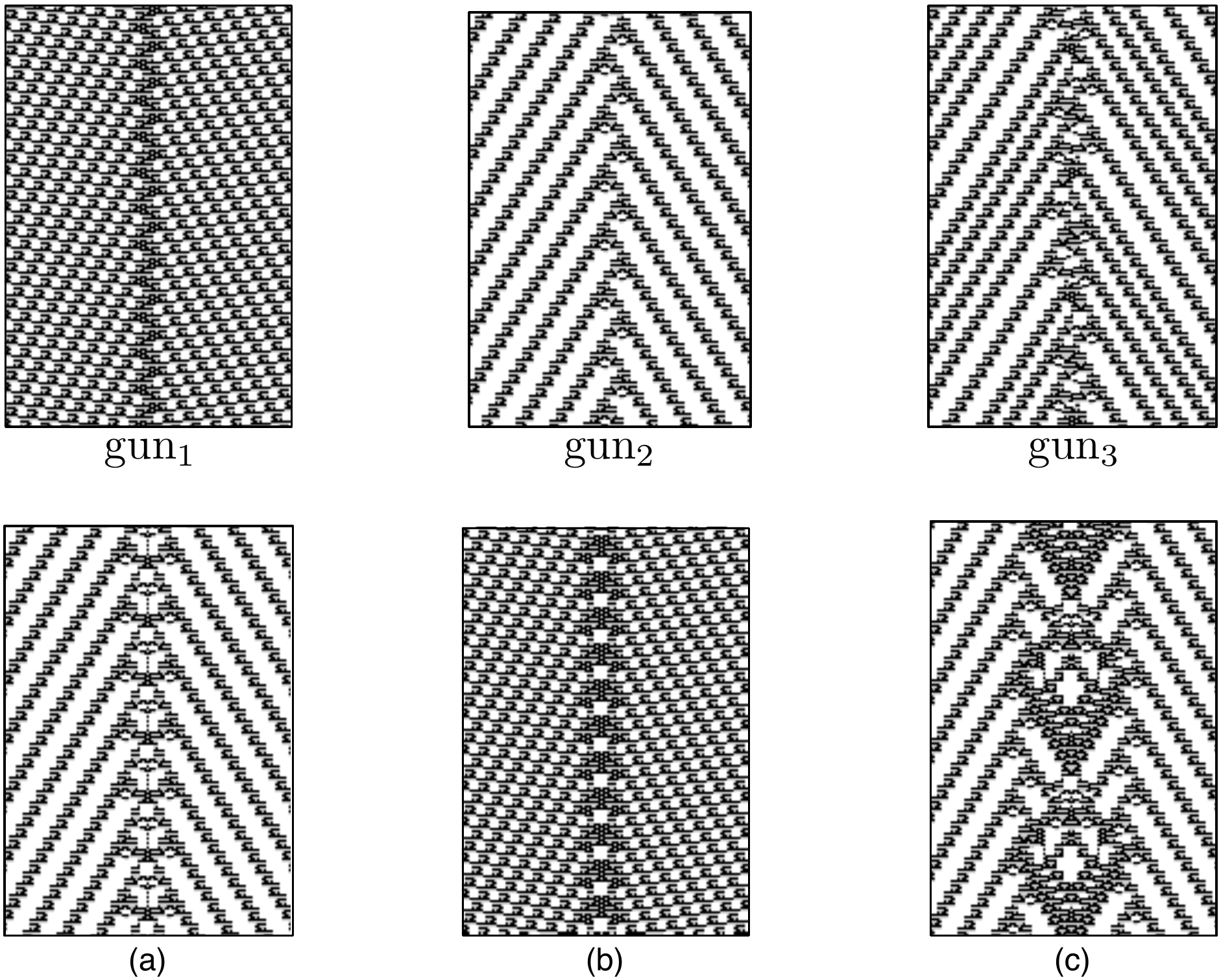}}
\caption{Samples of glider guns and their compositions in $\phi_{R126maj:4}$.}
\label{gliderGuns}
\end{figure}

Figure~\ref{gliderGuns} shows the three more frequent glider guns (gun$_1$, gun$_2$, and gun$_{3}$) emerging in $\phi_{R126maj:4}$. The next three guns (fig.~\ref{gliderGuns}(a), (b) and (c)) are combined in basic guns synchronized by multiple reactions which preserve  emission of gliders although some of them can get other frequencies. We can include them increasing the set of periodic structures $\cal{G}$$_{\phi_{R126maj:4}} = \{s_1, s_2, g_1, g_2, \mbox{gun}_1, \mbox{gun}_2,$ $ \mbox{gun}_3, \mbox{gun}_4\}$. The last gun is presented in fig.~\ref{selfOrgGun} as a collision of gliders.

Basic properties of $\cal{G}$$_{\phi_{R126maj:4}}$ are given in table~\ref{tableGliders}, where also glider guns can be classified by frequencies of glider emission as follows: 

\begin{quote}
\begin{itemize}
\item {\it small} $\mbox{gun}_1$ fires a gliders $g_1$ and $g_2$ (in turn) every five steps;
\item {\it medium} $\mbox{gun}_2$ fires $g_1$ and $g_2$ (in turn) every 26 steps;
\item {\it large} $\mbox{gun}_3$ fires five $g_1$ and $g_2$ gliders every 100 steps.
\end{itemize}
\end{quote}

Frequencies are related to the number of emitted gliders (in intervals) by a glider gun. Hence the gun$_2$ generates one $g_1$ and another $g_2$ glider eight steps before to produce the next ones, and a gun$_{3}$ yields five $g_1$ and five $g_2$ gliders 100 steps before to produce the other ones.

%%%%%%%%%%%%%%%%%%%%%%
\subsection{Collisions between gliders}

\begin{figure}[th]
\centerline{\includegraphics[width=4.3in]{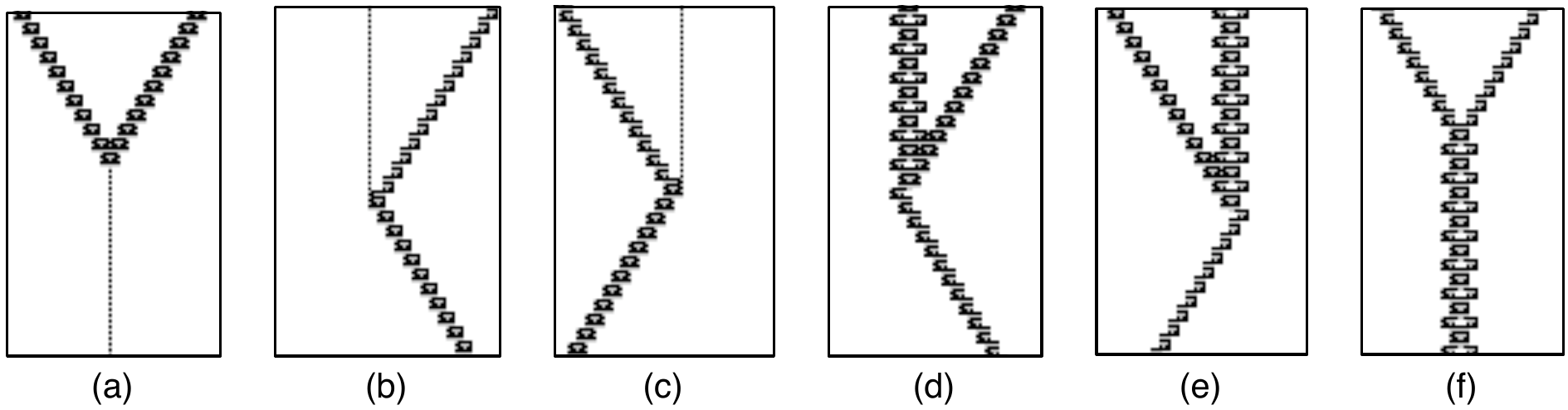}}
\caption{Generating basic gliders from collisions between other ones, this is self-organization by structures. The following reactions are illustrated as follows: (a) $g_1+g_2 = s_1$, (b) $s_1 + g_2 = g_1$, (c) $g_1 + s_1 = g_2$, (d) $s_2 + g_2 = g_1$, (e) $g_2 + s_1 = g_2$, and (f) $g_1 + g_2 = s_2$.}
\label{selfOrganization}
\end{figure}

Coding glider positions to get a desired reaction is a well-known solution for some related problems about complex behaviours. One of them is precisely the problem of self-organization (by structures fig.~\ref{selfOrganization}) \cite{kn:Kau93}. In this way, we present how each basic glider can be produced by collisions between other different gliders.

\begin{figure}[th]
\centerline{\includegraphics[width=4in]{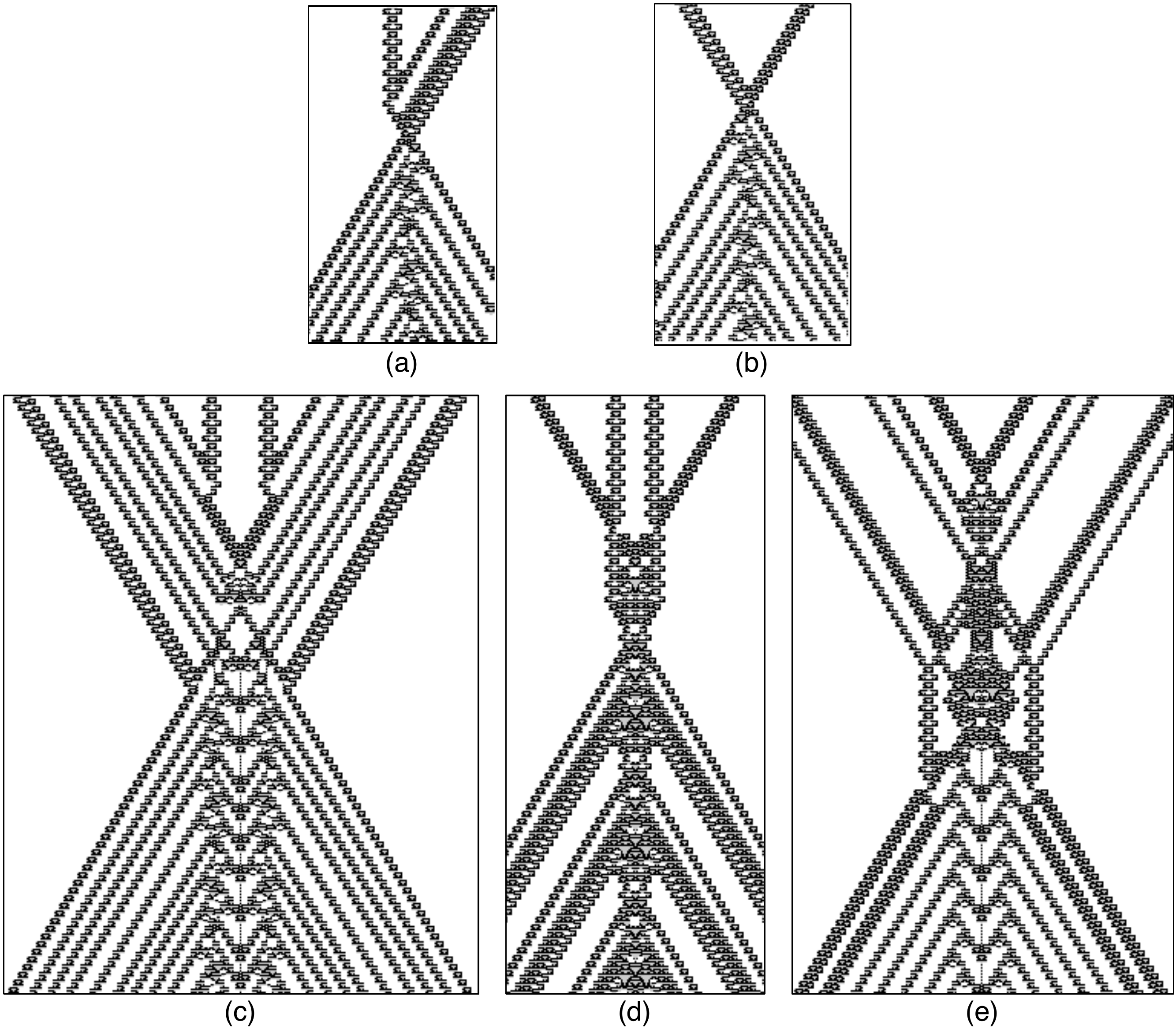}}
\caption{Generating gliders guns by collisions.}
\label{selfOrgGun}
\end{figure}

A little bit more complicated is to obtain glider guns by collisions. Figure~\ref{selfOrgGun}(a) and (b) depicts the production of a gun$_{2}$ from a multiple collision of gliders. Later on we shall present two new combinations of guns (fig.~\ref{selfOrgGun}(c) and (e)) and a new gun$_{4}$ (fig.~\ref{selfOrgGun}(d)).

%%%%%%%%%%%%%%%%%%%%%%
\subsection{Other collisions}

\begin{figure}[th]
\centerline{\includegraphics[width=3.7in]{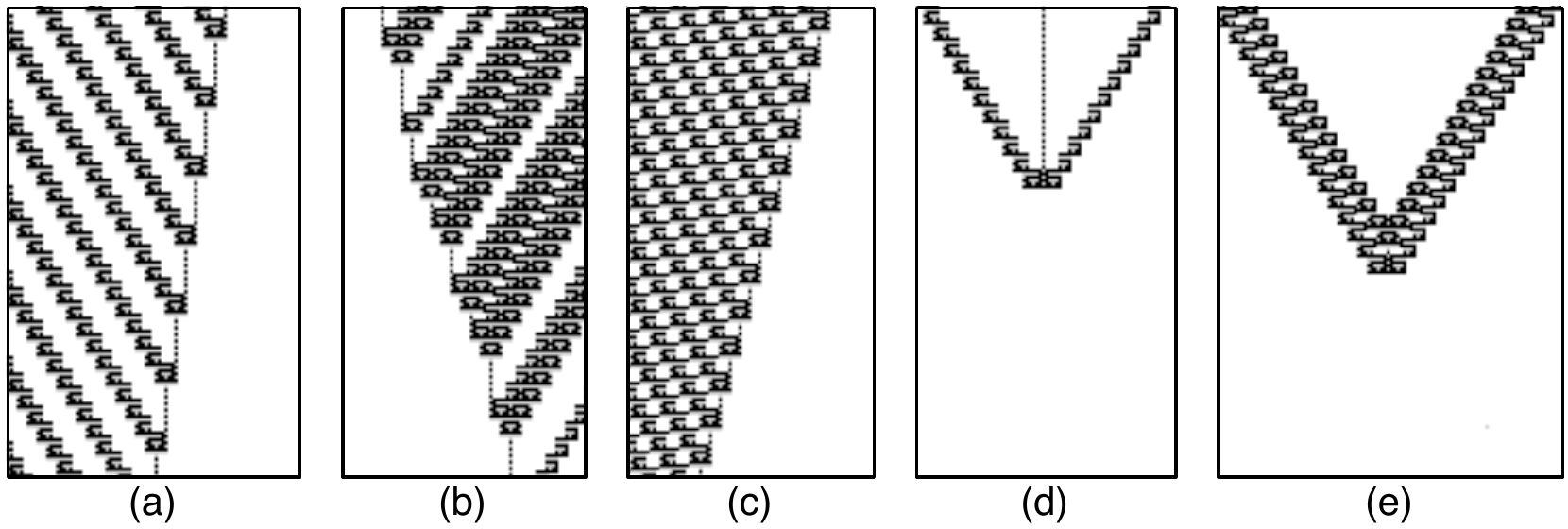}}
\caption{Eater reactions between gliders and still lives (a, b, and c) and annihilation of gliders by collisions (d and e).}
\label{eaters}
\end{figure}

A large variety of objects can be generated by collisions between gliders in $\phi_{R126maj:4}$. Some of these objects are useful for designing complex dynamical structures. For instance, fig.~\ref{eaters}(a), (b) and (c) shows how to delete (as an {\it eater} configuration) a single glider or a stream of gliders $g_1$ and $g_2$. Two kinds of annihilations are depicted as well (see fig.~\ref{eaters}(d) and (e)).

\begin{figure}[th]
\centerline{\includegraphics[width=4.8in]{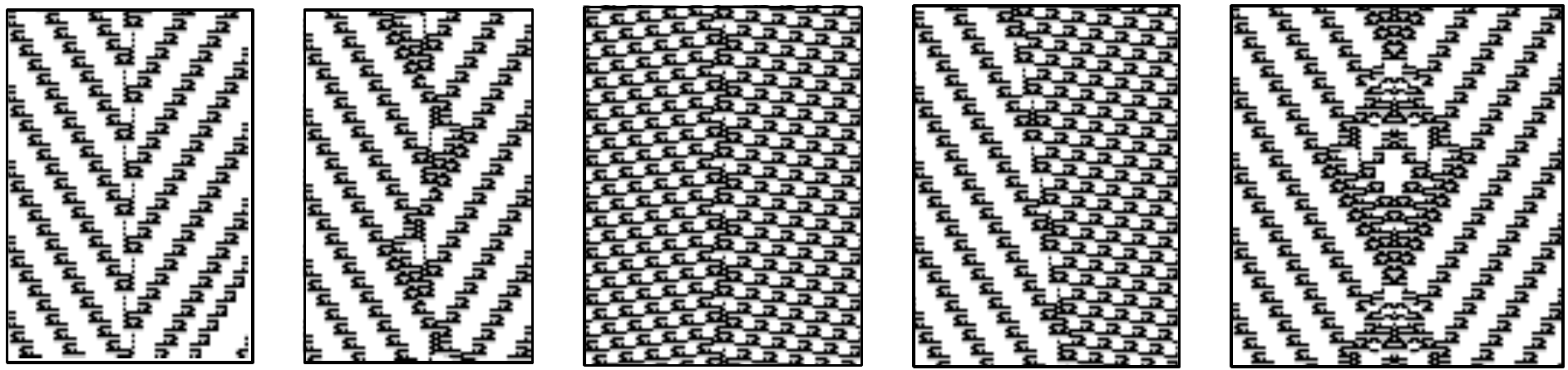}}
\caption{Black hole patterns consuming $g_1$ and $g_2$ gliders.}
\label{blackHole}
\end{figure}

\begin{figure}[th]
\centerline{\includegraphics[width=4.3in]{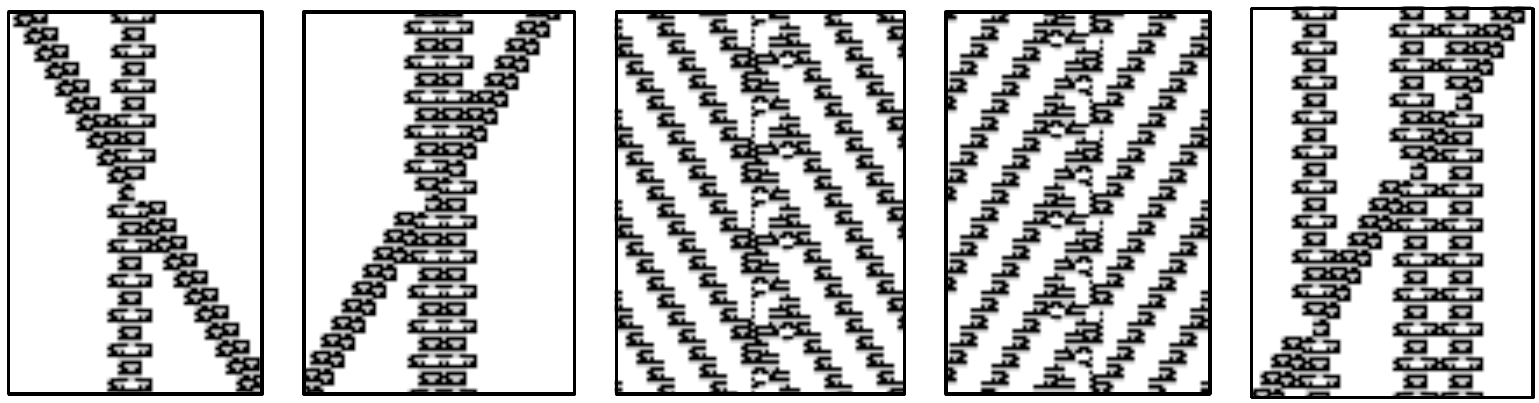}}
\caption{Examples of soliton reactions.}
\label{soliton}
\end{figure}

Figure~\ref{blackHole} describes a number of {\it black hole} patterns emerging in $\phi_{R126maj:4}$. Traditionally a black hole is a {\it Life} object\footnote{You can see a large classification of Life objects in \url{http://www.conwaylife.com/wiki/index.php}} absorbing any glider that comes close to the main body (in this case, still life patterns). Its relevance consists of knowing how many patterns are able of attracting gliders and consuming all of them forever, in these cases the  $g_1$ and $g_2$ gliders.

\begin{figure}[th]
\centerline{\includegraphics[width=4.8in]{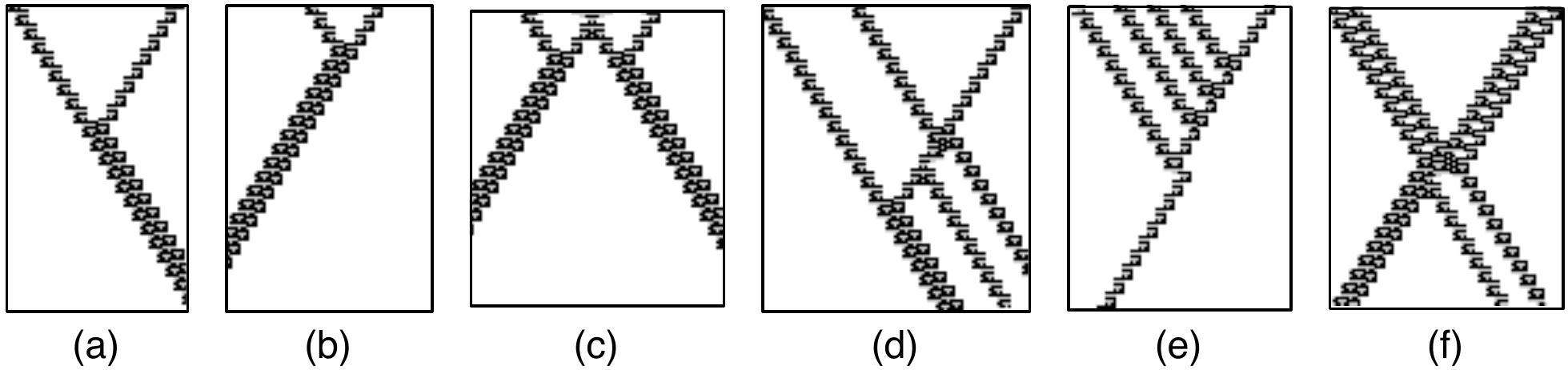}}
\caption{Examples of other collisions.}
\label{otherCollisions}
\end{figure}

Figure~\ref{soliton} displays soliton reactions, where colliding gliders preserve their original forms. As known, a soliton has a small change in their phase and displacement; also soliton reactions are obtained with single gliders or by a stream of them.

\begin{figure}%[th]
\centerline{\includegraphics[width=4.7in]{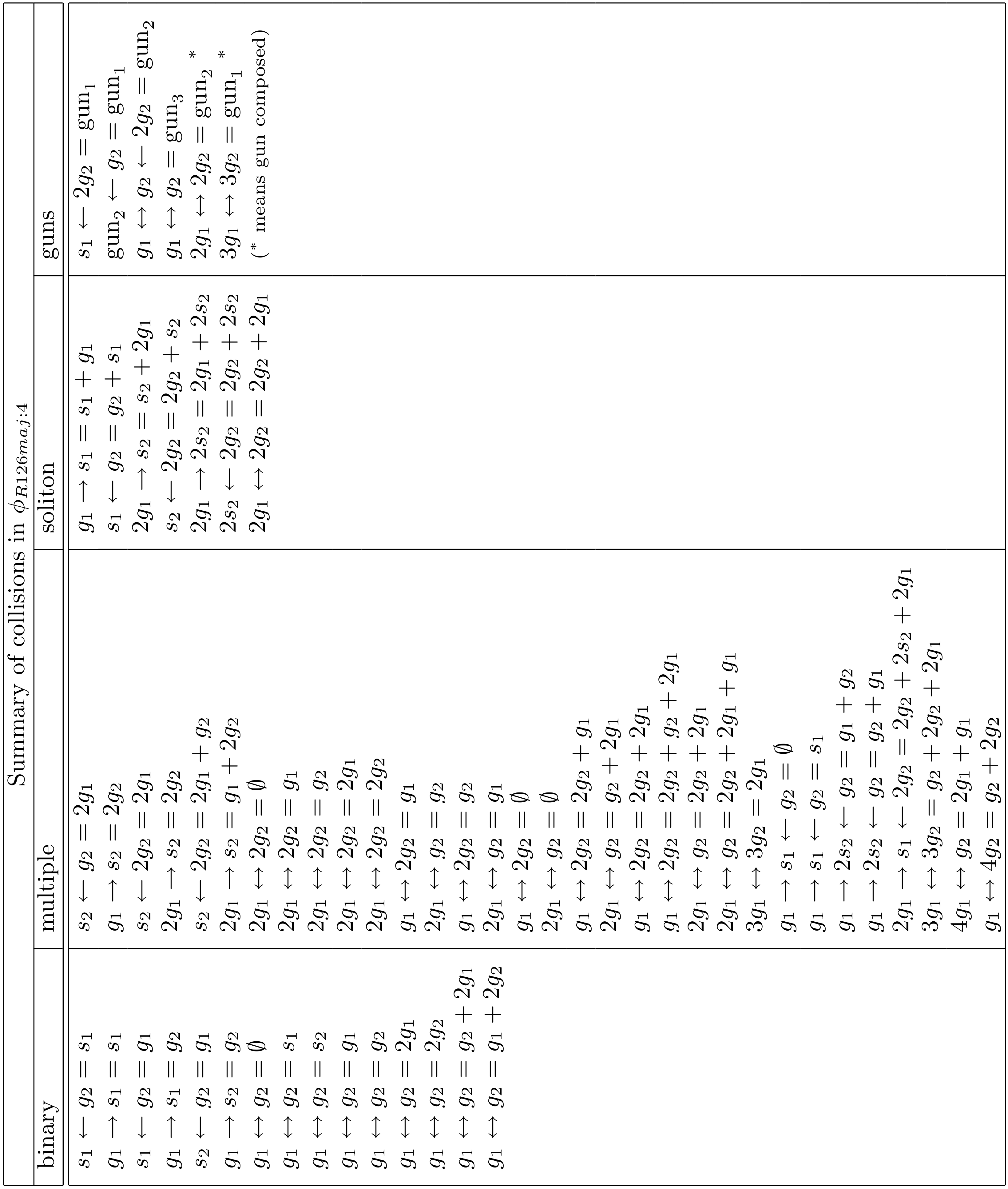}}
\caption{Table of binary, multiple and other collisions.}
\label{tableCollisions}
\end{figure}

The last set of examples (fig.~\ref{otherCollisions}) presents other binary reactions and some multiple collisions between $g_1$ and $g_2$ gliders. Some of them are conservative and others produce a new number of these structures. Finally we can take some collisions to exploit their dynamics and controlling glider reactions (see an extended relations of collisions in table~\ref{tableCollisions}). Thus it can be obtained a full number of reactions for generating desired gliders. The set of collisions is useful as `raw material' in the implementation of computations on $\phi_{R126maj:4}$.

%%%%%%%%%%%%%%%%%%%%%%
\section{Computing in $\phi_{R126maj:4}$}
Given the large number of reactions in $\phi_{R126maj:4}$ (see table~\ref{tableCollisions}) the rule could be useful for implementing collision-based computing schemes~\cite{kn:Ada03,kn:MAM06}.

\begin{figure}[th]
\centerline{\includegraphics[width=1.5in]{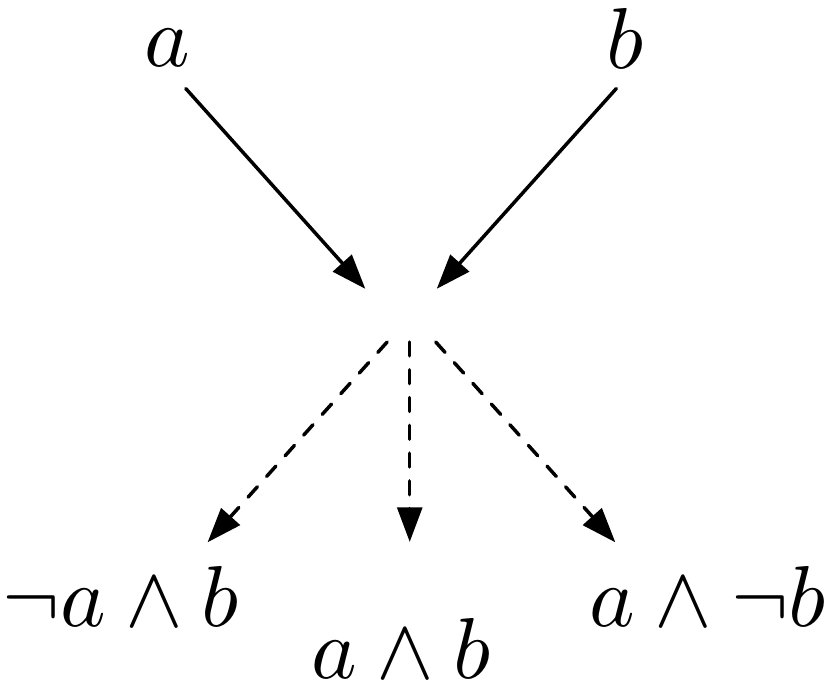}}
\caption{Colliding interactions deriving in logic gates.}
\label{gatesColliding-1}
\end{figure}

Figure~\ref{gatesColliding-1} illustrates the interaction of gliders traveling, colliding one another and implementing a Boolean conjunction as result. Initially from previous collisions we can embed logical constructions of {\sc and} and {\sc not} gates from this figure, as follows:

\begin{itemize}
\item for the gate ($\neg a \wedge b$) the implementation with $\phi_{R126maj:4}$ is in fig.~\ref{otherCollisions}(b); but only with one $g_2$ glider (see tab.~\ref{tableCollisions}).
\item for the gate ($a \wedge b$) the implementation corresponds to fig.~\ref{selfOrganization}(a).
\item for the gate ($a \wedge \neg b$) the implementation is presented in fig.~\ref{otherCollisions}(a); but only with one $g_1$ glider (see table~\ref{tableCollisions}).
\item for one {\sc fanout} gate ($a \leftrightarrow b = a + a + b$) the implementation is shown in fig.~\ref{otherCollisions}(d).
\end{itemize}

Indeed, here we also adopt ideas developed by Rennard in his design of Life computing architectures~\cite{kn:Ren03}. Glider $g_1$ represents  value 0, two $g_1$ gliders together represent a value 1. Two gliders $2g_2$ traveling in positive direction describe the operator and one the register. Thus the register will read {\sc false} or {\sc true} if they are produced successfully.

\begin{figure}[th]
\centerline{\includegraphics[width=4.5in]{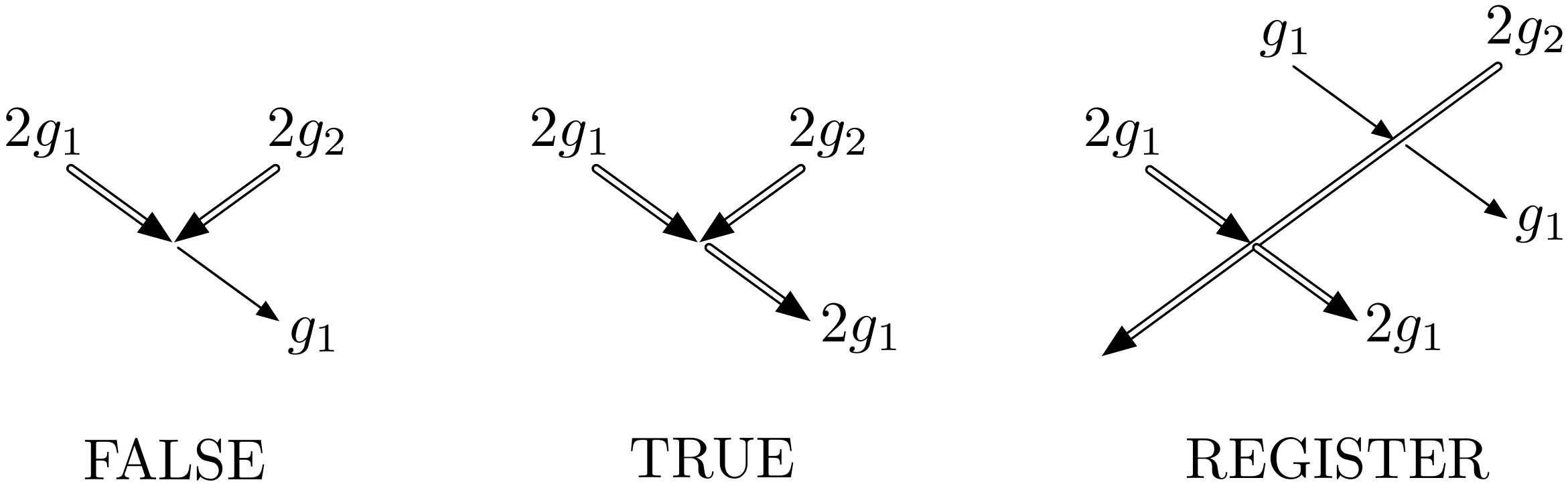}}
\caption{Example of devices working as data, operator and register respectively by gliders in $\cal G$$_{\phi_{R126maj:4}}$.}
\label{gatesColliding-2}
\end{figure}

Figure~\ref{gatesColliding-2} illustrates the basic reactions required to produce a primitive computational scheme in $\phi_{R126maj:4}$. The following set of relations is applied (see table~\ref{tableCollisions}):

\begin{center}
\small
\begin{tabular}{lll}
$2g_1 \leftrightarrow 2g_2 = \epsilon$ & $g_1 \leftrightarrow 2g_2 = g_1$ & $g_1 \leftrightarrow 2g_2 = 2g_2 + g_1$ \\
& $2g_1 \leftrightarrow 2g_2 = g_1$ & $2g_1 \leftrightarrow 2g_2 = 2g_2 + 2g_1$
\end{tabular}
\end{center}

\noindent so we can represent serial reactions as:

\begin{center}
\small
\begin{tabular}{ll}
$2g_1 + 2g_2 = \epsilon$ & empty word \\
$2g_1 + 2g_2 = g_1$ & {\sc false} \\
$2g_1 + 2g_2 = 2g_1$ & {\sc true}.
\end{tabular}
\end{center}

\noindent and a {\sc not} gate can be represented as:

\begin{itemize}
\item {\sc false}$ + 2g_2 =$ {\sc true}$+2g_2$, or
\item {\sc true}$ + 2g_2 =$ {\sc false}$+ 2g_2$.
\end{itemize}

%%%%%%%%%%%%%%%%%%%%%%
\subsection{Constructing formal languages since gliders collisions in $\phi_{R126maj:4}$}

To be considered as a mathematical machine, $\phi_{R126maj:4}$ should compute sets of formal languages~\cite{kn:Arb69}. We consider such implementation as an easy way to illustrate how implement some collision-based processes in $\phi_{R126maj:4}$.

Let $\Sigma$ be a non-empty finite alphabet. Thus a string over $\Sigma$ is a finite sequence of symbols from $\Sigma$. A set of all strings over $\Sigma$ of length $n$ is denoted by $\Sigma^n$. For example, if $\Sigma = \{0, 1\}$, then $\Sigma^2 = \{00, 01, 10, 11\}$ to $\Sigma^3 = \{000, 001, 010, 011, 100, 101, 110, 111\}$, and so on. This way $\Sigma^0 = \{\epsilon\}$ for any alphabet $\Sigma$. The set of all strings over $\Sigma$ of any length is the {\it Kleene closure} of $\Sigma$ and is denoted as $\Sigma^*$. However we can write such expression in terms of $\Sigma^n$, as follows:

\begin{equation}
\Sigma^* = \bigcup_{n \in Z_0} \Sigma^n.
\end{equation}

This way for a binary alphabet $\Sigma^*=\{\epsilon, 0, 1, 00, 01, 10, 11, 000, 001, \ldots\}$. Finally a set of strings on $\Sigma$ is called a formal language.

\begin{figure}[th]
\centerline{\includegraphics[width=3.5in]{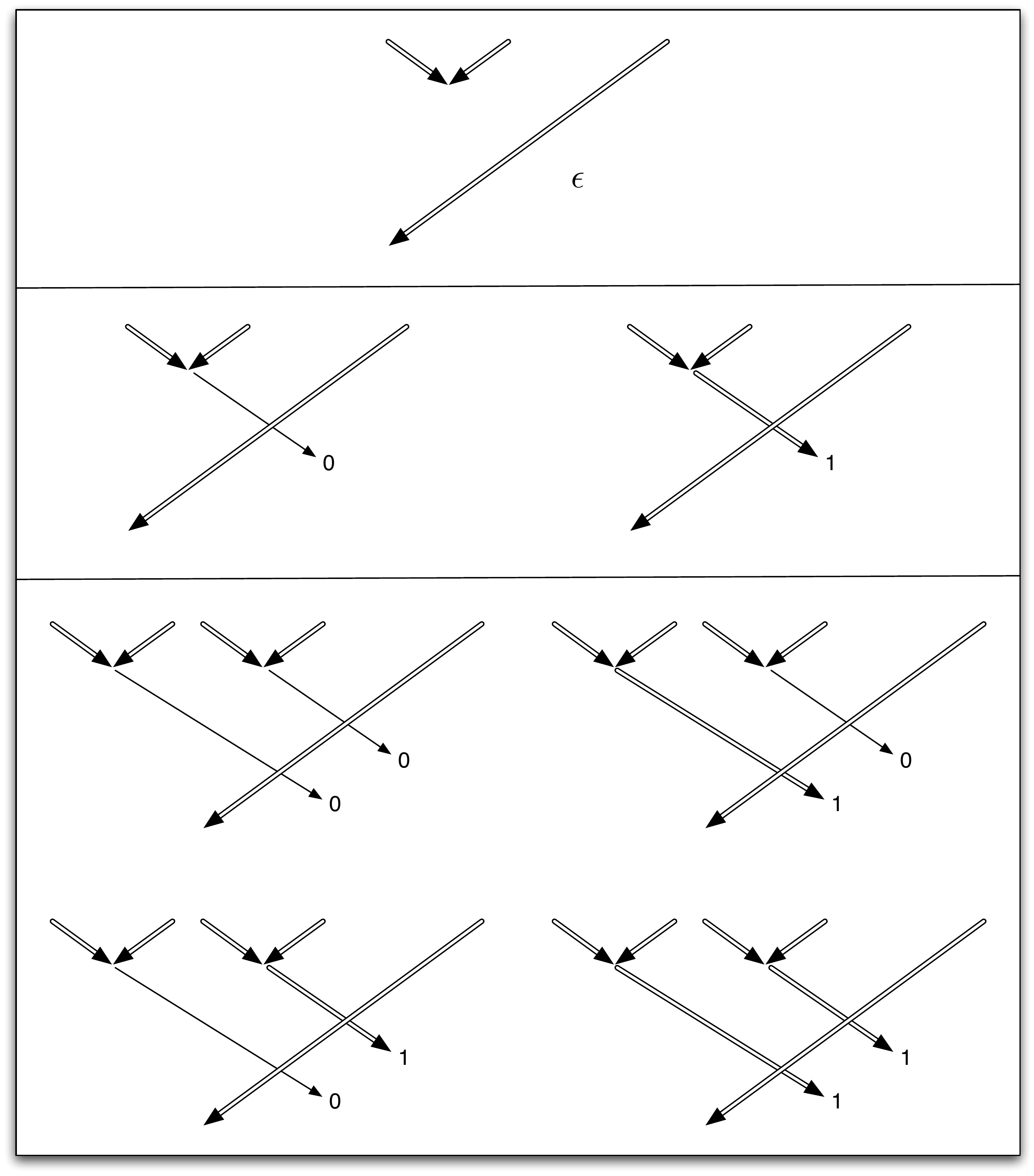}}
\caption{Constructing the formal collision-based languages in $\phi_{R126maj:4}$: They are $\Sigma^0$ (top), $\Sigma^1$ (middle), and $\Sigma^2$ (bottom).}
\label{gatesColliding-3}
\end{figure}

Making use of gliders in $\phi_{R126maj:4}$ we can get any set of strings of $\Sigma^*$. For example, for the formal language $1^*$ we need to code all the initial conditions with reactions {\sc true}. For the formal language $(00+11)^*$ the initial condition will be coded as: $(g_1 g_1 + 2g_1 2g_1)^*$. To yield an arbitrary length of strings we increase the number of cells in the CA. All distances between gliders must be preserved and the left part become periodic.

%%%%%%%%%%%%%%%%%%%%%%
\section{Conclusions}
We have enriched ECA Rule 126 with majority memory and have demonstrated that by applying certain filtering procedures we can extract rich dynamics of gliders, guns and infer a sophisticated system of reactions between gliders. We have discovered how a complex dynamics emerges from a chaotic system selecting an adequate memory. We have shown that the majority memory increases nominal complexity but decreases statistical complexity of patterns generated by the CA. By applying methods as de Brujin diagrams, cycles and graph theory, we have proved that Rule 126 with memory opens a new spectrum of complex rules, in this case a new CA with memory: $\phi_{R126maj:4}$. Finally we have demonstrated some capacities for computing specific logical and memory functions, and a future work will be constructing a universal device.

%%%%%%%%%%%%%%%%%%%%%%
\section*{Acknowledgement}
Genaro J. Mart\'{\i}nez and Ramon Alonso-Sanz are supported by EPSRC (grants EP/F054343/1 and EP/E049281/1). Juan C. Seck-Tuoh-Mora is supported by CONACYT (project CB-2007/83554).

%%%%%%%%%%%%%%%%%%%%%

\end{document}